\documentclass[11pt]{article}
\usepackage{amsmath,amssymb,color,graphicx,multirow,young}
\usepackage[flushleft]{threeparttable}
\usepackage[vcentermath]{youngtab}
\usepackage{hyperref}
\definecolor{darkred}{rgb}{0.65,0.15,0}
\hypersetup{pdfborder={0 0 0},colorlinks=true,urlcolor=darkred,citecolor=blue,linkcolor=darkred,linktocpage=true}
\usepackage{cite,amsfonts,amsthm,mathrsfs}
\usepackage{lscape}
\usepackage[T1]{fontenc}
\usepackage{enumerate}
\hypersetup{
	colorlinks   =  true,
	citecolor    = red,
	urlcolor	=magenta,
}
\def\4diml{four-dimensional}

\def\-1{^{-1}}

\makeatletter

\@addtoreset{equation}{section}
\makeatother
\begin{document}
	\thispagestyle{empty}
	\vspace{5mm}
	\begin{center}
		{\LARGE \bf  Non-Abelian T-duality of $AdS_{d\le3}$ families by \\[2mm] Poisson-Lie T-duality }

		\vspace{14mm}
		\normalsize
		{\large  Ali Eghbali\footnote{eghbali978@gmail.com}, Reza Naderi\footnote{r\_naderi8@yahoo.com},
		Adel  Rezaei-Aghdam\footnote{rezaei-a@azaruniv.ac.ir}}\\
	
\vspace{4mm}
{\small {\em Department of Physics, Faculty of Basic Sciences,\\
		Azarbaijan Shahid Madani University, 53714-161, Tabriz, Iran}}\\
\vspace{10mm}
\begin{tabular}{p{14cm}}
{\small
We proceed to investigate the non-Abelian T-duality of $AdS_{2}$, $AdS_{2}\times S^1$
and $AdS_{3}$ physical backgrounds, as well as the metric of the analytic continuation of $AdS_{2}$ from the point of view of  Poisson-Lie (PL) T-duality.
To this end, we reconstruct these metrics of the $AdS$ families as backgrounds of non-linear $\sigma$-models on two- and three-dimensional Lie groups.
By considering the Killing vectors of these metrics and by taking into account the fact that the subgroups of isometry Lie group of
the metrics can be taken as one of the subgroups of the Drinfeld double (with Abelian duals) we look up the PL T-duality.
To construct the dualizable metrics by the PL T-duality we find all subalgebras of Killing vectors that generate subgroup of isometries which acts freely and transitively on the manifolds defined by aforementioned $AdS$ families.
We then obtain the dual backgrounds for these families of $AdS$ in such a way that we apply the usual rules of PL
T-duality without further corrections. We have also investigated the conformal invariance conditions of the original backgrounds ($AdS$ families)
and  their dual counterparts. Finally, by using the T-duality rules proposed by Kaloper and Meissner (KM)
we calculate the Abelian T-duals of BTZ black hole up to two-loop by dualizing on the coordinates $ \varphi$ and $ t $.
When the dualizing is implemented by the shift of direction $\varphi$,
we show that the horizons and singularity of the dual spacetime are the same as in charged black string derived by Horne and Horowitz without $\alpha'$-corrections,
whereas in dualizing on the coordinate $t$ we find a
new three-dimensional black string whose structure and asymptotic nature are clearly determined.
For this case, we show that the T-duality transformation changes
the asymptotic behavior from $AdS_3$ to flat.}

		\end{tabular}
		\vspace{6mm}
	\end{center}	
	\newpage
	\setcounter{page}{1}
	\tableofcontents
	%%%%%%%%%%%%%%%%%%%%%%%%%%%%%%%%%%%%%%%%%%%%%%%%%%%%%%%%%%%%%%%%%%%%%%%%%%%%%
	\vspace{5mm}
	\vspace{5mm}
	\section{Introduction}
	\label{Sec.I}
Anti-de Sitter (AdS) space as a constant-negative-curvature spacetime is the maximally
symmetric solution of Einstein's equation with a negative cosmological constant. $AdS_d$ space
can be realized as a hyperboloid embedded in a $d + 1$-dimensional geometry whose metric may be,
in $d$-dimensional half-space coordinate, expressed as
\begin{eqnarray}\label{1}
ds^2&=&\frac{l^2}{y^2}(-dt^2+dy^2+\sum^{d-2}_{i=1}dx_i^2), \end{eqnarray}
where $l$ is the radius of the $AdS_d$ metric. In this
work we will deal with  $AdS_{2}$,  $AdS_{2}\times S^1$ and $AdS_{3}$ backgrounds, and also the metric of the analytic continuation of $AdS_{2}$ in Poincar\'{e} coordinates.
The $AdS_{3}$ metric is, in  Poincar\'{e} coordinates, given by
\begin{eqnarray}\label{2}
ds^2 = l^2 (\frac{dr^2}{r^2}-r^2 dt^+ dt^-).
\end{eqnarray}
This metric is an exact solution of
Einstein's equation so that it covers a subregion of $AdS_{3}$ called the Poincar\'{e} patch.
A spacetime is called asymptotically $AdS$ if it approaches \eqref{2} as $r\rightarrow \infty$.
$AdS$ backgrounds have made numerous appearances in the context of the AdS/CFT correspondence, as well as the
string theory.
An example of an asymptotically $AdS$ spacetime is the BTZ black hole.
The BTZ  as a $2+1$-dimensional black hole solution with mass, charge, angular momentum, and negative cosmological constant was, first, found by Banados, Teitelboim and  Zanelli  \cite{Banados}.
A slight modification of this black hole solution
yields an exact solution to string theory \cite{Horowitz2}.
The BTZ black hole as an $AdS_3$ spacetime is a solution for the equations of motion of the low energy string effective action with
antisymmetric tensor field $B_{\varphi t} =r^2/l$ and a zero dilaton field \cite{Horowitz2}. By studying the dual of this solution \cite{Horowitz2}, it was shown that the BTZ black hole solution is, under the Abelian T-duality, equivalent to
the charged black string solution discussed in \cite{Horowitz1}.
Abelian T-duality \cite{Buscher1} is a well known symmetry of string theory that maps any
solution of the string equations with a
translational symmetry to another solution. In fact, the duality symmetry is one of the most interesting properties of string theory
that different spacetime geometries can correspond to
equivalent classical solutions.
In Ref. \cite{Tseytlin1}, it was discussed the Abelian T-duality for $AdS_2$, as well as the two-sphere $S^2$.
These two examples are different in nature as T-duality performed along a compact direction for $S^2$ and along a non-compact direction for
$AdS_2$.
In the case of the non-Abelian T-duality of $AdS_3$ spacetime, it was shown that \cite{EMR13} the non-Abelian T-duality transformation (as the PL T-duality on a semi-Abelian double) relates the $AdS_3$ spacetime with no horizon and no curvature singularity (the BTZ vacuum solution) to
a solution with a single horizon and a curvature singularity (the charged black string).
The PL T-duality is a generalization of Abelian \cite{Buscher1} and non-Abelian dualities
\cite{nonabelian} that proposed by  Klimcik and Severa \cite{Klim1}.
They showed that the dual $\sigma$-models can be formulated on the Drinfeld double group $\bf D$ \cite{Drinfeld}, which by definition has a pair of
maximally isotropic subgroups $\bf G$ and ${\tilde {\bf G}}$ corresponding to the subalgebras ${\cal G}$ and ${\tilde {\cal G}}$,
such that the subalgebras are duals of each other in the usual sense, i.e.,
${\tilde {\cal G}} = {\cal G}^\ast$.
As mentioned above, the $AdS_3$ background is a solution for the low energy string effective action equations
with a nonzero field strength $H=dB$. In the present work, we consider the $AdS_{3}$ background with zero B-field, i.e. $H=0$.
Indeed, an argument \cite{Horowitz3} claiming that $AdS_{3}$ solutions to the low energy string equations do not exist assumed that $H=0$.
Obviously, in the case of two-dimensional backgrounds such as $AdS_{2}$,  the field strength $H$ is always absent.

The main goal of this paper is to investigate the non-Abelian T-duality of $AdS_{2}$, $AdS_{2}\times S^1$
and $AdS_{3}$ backgrounds, as well as the metric of the analytic continuation of $AdS_{2}$ by using PL T-duality approach. For that purpose,
we first obtain the isometry subgroups of these metrics acting
freely and transitively on the corresponding target space manifolds.
As we will address in the following, these isometry subgroups can be used
for construction of their non-Abelian T-dual backgrounds. Sufficient condition for that is
that the metrics have an isometry subgroup whose dimension is equal to
the dimension of the manifold and its action on the manifold is free and transitive.
This procedure was first applied for homogenous plane-parallel wave metric in \cite{Hlavaty1} (see, also, \cite{Hlavaty11}).
On the other hand, we have lately studied the Abelian T-duality of G\"{o}del string cosmologies up to $\alpha'$-corrections \cite{Godel} by
applying the T-duality rules at two-loop order which were obtained by KM in \cite{KM}.
Using the T-duality rules of KM, we shall study the Abelian T-duality of BTZ black hole up to $\alpha'$-corrections
by dualizing on both directions $\varphi$ and $t$.
When the dualizing is implemented by the shift of direction $\varphi$ we show that
the structure and  asymptotic nature of the dual
spacetime including the horizons and singularity are the same as in charged black string derived by Horne and Horowitz \cite{Horowitz1} without $\alpha'$-corrections,
whereas in performing the duality with respect to the coordinate $t$ we find a
new three-dimensional black string whose structure and asymptotic nature including the horizons and singularity are also determined.
Moreover, we show that the Abelian T-duality transformation of KM changes
the asymptotic behavior of solutions from $AdS_3$ to flat.
In the case of our black string, it is interesting to note that the true singularity lies outside the horizons.
Similar to this, by using the version of the general relativity
field equations produced in \cite{suggett1} it has been shown that \cite{suggett2} there is a singularity outside of a Schwarzschild black hole.
Perhaps the most important feature of our black string solution is that its global structure is
qualitatively different from previously discussed string solution, where dualizing is performed by the shift of direction $\varphi$.
%%%%%%%%%%%%%%%%%%%%%%%%%%%%%%%%%%%%%%%%%%%%%%%%%%%%%%%%%%%%%%%%

This paper is organized as follows. After Introduction section,  Sec. \ref{Sec.II} reviews the construction of
PL T-dual $\sigma$-models over the Lie groups, where necessary formulas are summarized. We furthermore review
the conformal invariance conditions of the bosonic $\sigma$-model up to two-loop order at the end of this section.
We start Sec. \ref{Sec.III} by investigating the conformal invariance conditions of the $AdS_2$ metric up to $\alpha'^2$-corrections (three-loop order).
Then, the non-Abelian target space dual of the metric is obtained by using the Lie subgroup of isometry group acting
freely and transitively on the $AdS_{2}$ manifold; we also give a note on the non-Abelian T-duality of the metric of the analytic continuation of $AdS_{2}$
at the end of this section.
Similar to the construction of T-dual $\sigma$-model for $AdS_2$ in Sec. \ref{Sec.III}, we obtain the non-Abelian T-dual background of
the $AdS_{2}\times S^1$ metric in Sec. \ref{Sec.IV}; the results of this section are summarized in Tables \ref{table:1} and \ref{table:2}.
The study of the non-Abelian T-dualization of the $AdS_3$ metric is given in Sec. \ref{Sec.V}.
The results of this section including the constant matrix $E_0(e)$,
the transformation between $ AdS_3 $ coordinates and group ones, and
the metrics and $B$-fields corresponding to both original and dual backgrounds are clearly displayed
in Table \ref{table:4}.
The study of Abelian T-duality of BTZ black hole up to $\alpha'$-corrections by
using the KM approach, when the duality is implemented by a shift of the coordinates $\varphi$ and $t$,
is discussed in Sec. \ref{Sec.VI}.
Some concluding remarks are given in the last section.

%%%%%%%%%%%%%%%%%%%%%%%%%%%%%%%%%%%%%%%%%%%%%%%%%%%%%%%%%%%%%%%%%%%%%%%%%%%%%

\section{Some review on PL T-duality and two-loop conformal invariance}
\label{Sec.II}
In this section, we begin by reviewing the construction of PL T-dual $\sigma$-models over the Lie groups.
In order to investigate the conformal invariance conditions of the models constructed out by
the metrics $AdS_2 $, $AdS_{2}\times S^1$, $ AdS_3 $, and also the metric of the analytic continuation of $AdS_{2}$
up to the first order in  $\alpha'$ we also review the two-loop beta function equations proposed by Hull and Townsend (HT) \cite{c.hull}.

\subsection{A brief review of PL T-duality}
\label{Sec.II.1}
Here we review non-Abelian T-duality via the PL T-duality approach in the absence of spectators. According to \cite{Klim1} the PL duality is based on the concept of Drinfeld double\cite{Drinfeld}. The Drinfeld double $\bf D$ is a Lie group whose Lie algebra $ \mathcal{D} $ as a vector space can be decomposed into direct sum of two Lie subalgebras  $ \mathcal{G} $ and $ \tilde {\mathcal{G}} $, such that $ \mathcal{D} $ is maximal isotropic with respect to a non-degenerate invariant bilinear form $ \langle  ~,~\rangle $ on  $ \mathcal{D} $. Actually, by taking the sets $ \{T_a\} $ and  $ \{\tilde T^a\} $ as the bases of the Lie algebras   $ \mathcal{G} $ and $ \tilde {\mathcal{G}} $, respectively, we have:
		\begin{eqnarray}
			[T_a , T_b] &=& f_{ab}^{~c} ~T_c,~~~\label{DC}
			[{\tilde T}^a , {\tilde T}^b] = {{\tilde f}^{ab}}_{\; \; \: c} ~{\tilde T}^c,\nonumber\\
			{[T_a , {\tilde T}^b]} &=& {{{\tilde f}^{bc}}_{\; \; \; \:a} {T}_c + {f_{ca}^{~b}} ~{\tilde T}^c},
		\end{eqnarray}
	where $ f_{ab}^{~c} $ and $ {{\tilde f}^{ab}}_{\; \; \: c} $ are structure constants of the Lie algebras $\mathcal{G} $ and $ \tilde {\mathcal{G}} $, respectively.
	The isotropy of the subalgebras with respect to bilinear form means that
			\begin{eqnarray}
		 \langle  T_a,T_b\rangle=\langle  {\tilde T}^a,{\tilde T}^b\rangle=0,~~\langle  T_a,{\tilde T}^b\rangle=\delta_a^{~b} .\label{bil}
	\end{eqnarray}
	Also, the Jacobi identity of Lie algebra $ \mathcal{D} $ imposes the following mixed Jacobi relations over the structure constants of Lie algebras    $ \mathcal{G} $ and $ \tilde {\mathcal{G}}  $
	\begin{eqnarray}
		{f^a}_{bc}{\tilde{f}^{de}}_{\; \; \; \; a}=
		{f^d}_{ac}{\tilde{f}^{ae}}_{\; \; \; \;  b} +
		{f^e}_{ba}{\tilde{f}^{da}}_{\; \; \; \;  c}+
		{f^d}_{ba}{\tilde{f}^{ae}}_{\; \; \; \; c}+
		{f^e}_{ac}{\tilde{f}^{da}}_{\; \; \; \; b}.
	\end{eqnarray}
In order to define $ \sigma $-models with PL duality symmetry, we need to consider the following relations \cite{Klim1}:
	\begin{eqnarray}\label{abPi}
		  g^{-1} T_{c}~g = a(g)_{c}^{~b} ~ T_{b},&
		g^{-1} {\tilde T}^{a} g = b(g)^{ac} ~ T_{c}+d(g)^a_{~c} {\tilde T}^{c},&
{\varPi}(g)= b(g)a^{-1}(g),
	\end{eqnarray}
where $g$ is an element of the Lie group $\bf G$ corresponding to the Lie algebra $\mathcal{G}$.
 The invariance of inner product with respect to adjoint action of group together with \eqref{bil} and \eqref{abPi} requires the above matrices to possess the following properties\cite{{Klim1},{Sfetsos}}\footnote{Here ``t'' denotes transposition.}:
	\begin{eqnarray}\label{2.5}
d(g)=a^{-t}(g),~~~a^{-1}(g)=a(g^{-1}),~~~b^t(g)=b(g^{-1}),~~~\varPi^t(g)=-\varPi(g).
\end{eqnarray}
Now, one may define below the $ \sigma $-model with $d$-dimensional target manifold $\cal M$ where the action of group $\bf G$ on $\cal M$ is free and transitive so that ${\cal M} \approx {\bf G}$ \cite{Klim1}
\begin{eqnarray}
S=\frac{1}{2}\int_{\Sigma} d\sigma^+d\sigma^- ~ {R^a_{+}} {R^b_{-}} E_{ab}(g), \label{Sigma1}
\end{eqnarray}
where $\sigma^{\pm}$ stand for the standard light-cone variables which are defined by means of the coordinates of worldsheet ${\Sigma}$, giving
$\sigma^{\pm} =(\tau \pm \sigma)/2$ together with $\partial_{\pm}=\partial_{\tau} \pm \partial_{\sigma}$,
and ${R^a_{\pm}}$ are the components of right invariant one-forms on $ \bf G $ which are defined in the following way
\begin{eqnarray}
{R^a_{\pm}} = (\partial_\pm g g^{-1})^a = \partial_{\pm}X^{^M}~ {R_{_M}}{^a}, \label{RIOF}
\end{eqnarray}
where $X^{^{M}}, {M} =1,...,d$ are the coordinates of manifold $\cal M$. The background matrix $E$ is given in matrix notation by
\begin{eqnarray}
	E(g)=\big(E_0^{-1}(e)+\varPi(g)\big)^{-1},\label{2.8.1}
\end{eqnarray}
where $ E_0 (e) $ is an invertible constant matrix in which $e$ is the unit element of the group ${\bf G}$.
The target space of dual model is the $d$-dimensional manifold $ \tilde {\cal M} $ with
coordinates ${\tilde X}^{^{M}}$.
Analogously, the group $ \tilde {\bf G}$ (whose dimension is, however,
equal to that of $G$) corresponding to $\tilde {\cal G}$ acts freely and transitively on the manifold $ \tilde {\cal M}$
so that $\tilde {\cal M} \approx {\tilde {\bf G}}$.
Then, the corresponding dual action can be written as \cite{Klim1}
	\begin{eqnarray}
	\tilde S=\frac{1}{2}\int_{\Sigma} d\sigma^+d\sigma^-~{\tilde R}_{+_a} {\tilde R}_{-_b} ~{\tilde E^{ab}} (\tilde g),\label{Sigmat1}
\end{eqnarray}
where $\tilde g$ is an element of the Lie group $\tilde {\bf G}$,
and ${\tilde R}_{\pm_a}$  are the components of right invariant one-forms on ${\tilde {\bf G}} $ which are defined as in \eqref{RIOF}.
The background appearing in this action are given in matrix notation by
\begin{eqnarray}
{\tilde E}(\tilde g)=\big(E_0(e)+\tilde{\varPi}(\tilde g)\big)^{-1},\label{Et}
\end{eqnarray}
where $\tilde{\varPi}(\tilde g)$ is defined as in \eqref{2.5} by replacing untilded quantities by
tilded ones and vice versa.
It should be noted that the actions \eqref{Sigma1} and \eqref{Sigmat1} correspond to PL dual $\sigma$-models \cite{Klim1}.
If the group $ \bf G$ becomes the isometry group of the manifold $\cal M$ with dual Abelian group $ \tilde {\bf G} $, then one can obtain the standard non-Abelian duality \cite{nonabelian}.

Let us now compare the $\sigma$-model \eqref{Sigma1} with the following
standard two-dimensional non-linear $\sigma$-model defining on the worldsheet in $d$ spacetime dimensions
\begin{eqnarray}
S_{_{nlsm}} =\frac{1}{2}\int_{{\Sigma}} d\sigma^+d\sigma^-  \mathcal{E}_{_{MN}}(X)  \partial_+X^{^M} \partial_-X^{^N}, \label{Sigma2}
\end{eqnarray}
where the background $\mathcal{E}_{_{MN}} ={G}_{_{MN}} + {B}_{_{MN}}$ combines the metric
${G}_{_{MN}}$ on $\cal M$ with antisymmetric tensor field ${B}_{_{MN}}$ ($B$-field).
The fields $X^{^{M}}(\sigma^{\alpha})$ constitute a map from the worldsheet to the manifold $\cal M$
with local coordinates $X^{^{M}}$.
Comparing \eqref{Sigma1} and \eqref{Sigma2} and then using \eqref{RIOF} one obtains that
\begin{eqnarray}\label{2.12}
\mathcal{E}_{_{MN}}  =  {R_{_M}}{^a} {R_{_N}}{^b} E_{ab}(g).
\end{eqnarray}
Notice that for the non-Abelian duality case, $\tilde f^{ab}_{~~~c}=0$, we find that $ b(g)=0 $, then, $ \varPi(g)=0 $; consequently,
$ E(g)=E_0(e)$. In this case, if $E_0(e)$ is chosen to be symmetric, then one concludes that the $B$-field  vanishes.
In general $E_0(e)$ of \eqref{2.8.1} can have an antisymmetric part, and in that case the $B$-field would be non-vanishing.

Before closing this subsection, we shall give a short summary of the dualization procedure that we will apply in the next sections.
First of all, we obtain the subalgebras of $d(d+1)/2$-dimensional isometry Lie algebras that generate subgroups
of isometry Lie groups that act freely and transitively on $d$-dimensional target space $\cal M$ where
the metrics of $AdS$ families are defined.
Since the dualizable metrics can be constructed by virtue of
Drinfeld double, the isometry subgroups of the metrics are taken as one of the subgroups of the Drinfeld double. In order to satisfy the dualizability conditions the other subgroup is considered to be Abelian. In other words, since we are dealing with non-Abelian T-duality,
the dual Lie algebra must be chosen Abelian.
In order to obtain the metrics of $AdS$ families by the Drinfeld double construction
we need to find the transformation between
group (isometry subgroup) coordinates and geometrical coordinates by choosing a convenient element of group.
In this case, we have to get the left-invariant vector fields on the group.
Accordingly, the metrics can be transformed into the group coordinates and finally one can write the corresponding actions to the transformed metrics in the from
of \eqref{Sigma2}. Thus, one may use \eqref{2.12} to obtain the original $\sigma$-model from \eqref{Sigmat1} and then dual one from
\eqref{Sigma1}.

%%%%%%%%%%%%%%%%%%%%%%%%%%%%%%%%%%%%%%%%%%%%%%%%%%%%%%%%%%%%%%%%%%%%%
\subsection{Two-loop conformal invariance conditions of the bosonic string $\sigma$-model}
\label{Sec.II.2}
In order to study the conformal invariance conditions of the bosonic string $\sigma$-model,
Fradkin and Tseytlin \cite{FRADKIN} have suggested that one should add to action \eqref{Sigma2} the renormalizable, but not Weyl invariant, term
\begin{eqnarray}\label{a.1}
	S_{_{dil}} = \frac{1}{8}\int_{{\Sigma}} d\sigma^+d\sigma^-  R^{^{(2)}}  \Phi(X),
\end{eqnarray}
where $R^{^{(2)}}$ is the curvature scalar for the worldsheet metric, and $\Phi(X)$
is the background dilaton field in the manifold $\cal M$.
The term \eqref{a.1} breaks Weyl invariance on a classical level as do the one-loop corrections to ${G}$ and  ${B}$.
It is essential for string consistency that, as a quantum field theory, the $\sigma$-model
be locally scale invariant.  This is equivalent to the requirement that the two-dimensional
worldsheet stress-energy tensor of the theory be traceless. In the $\sigma$-model $S_{_{nlsm}}  + S_{_{dil}}$,
local scale invariance is broken explicitly by the term \eqref{a.1}.
Consistency of the string theory requires that the action $S_{_{nlsm}}  + S_{_{dil}}$
defines a conformally invariant quantum field theory.
In the $\sigma$-model context, the conformal invariance conditions of the $\sigma$-model $S_{_{nlsm}}  + S_{_{dil}}$ are
provided by the vanishing of the three functions ${\beta_{_{MN}}^G}$, ${\beta_{_{MN}}^B}$ and
${\beta^\Phi}$ (beta function equations) \cite{callan}. In the HT scheme,
these equations are, at the two-loop level (first order in $\alpha'$)
\footnote{The dimensionful coupling constant $\alpha'$ turns out to be the inverse string tension.},
given by \cite{{c.hull},{Metsaev}} \footnote{We note that round brackets denote the symmetric part
on the indicated indices whereas square brackets denote the antisymmetric part.}
\begin{eqnarray}
{\beta_{_{MN}}^G}&=& {{\beta_{_{MN}}^G}}^{\hspace{-2mm}(1)} +\alpha' ~ {{\beta_{_{MN}}^G}}^{\hspace{-2mm}(2)}+{\cal O}(\alpha'^2) \nonumber\\
	&=&{\cal R}_{_{MN}}-\frac{1}{4}H^2_{_{MN}}+2{\nabla}_{_M}
	{\nabla}_{_N} \Phi +\frac{1}{2} \alpha' \Big[{\cal R}_{_{M PQR}} {\cal R}_{_N}^{^{~PQR}}
	+\frac{1}{2} {\cal R}_{_{M PQ N}} {H^2}^{^{PQ}}\nonumber\\
	&&~~~~~~~+\frac{1}{2} {\cal R}_{_{PQ R(M}}H_{_N)}^{^{~R S}} H^{^{PQ}}_{~~_{S}} +\frac{1}{12} ({\nabla}_{_M} H_{_{PQR}})
	({\nabla}_{_N} H^{^{PQR}})-\frac{1}{4}({\nabla}_{_P} H_{_{RS M}})
	({\nabla}^{^P} H^{^{RS}}_{~~_{N}})\nonumber\\
	&&~~~~~~~+\frac{1}{8} H_{_{MPQ}} H_{_{N R S}} H^{^{T S Q}} H_{_T}^{^{~~R P}}
	+\frac{1}{8} H_{_{M P Q}} H_{_{NR}} ^{^{~~Q}}   {H^2}^{^{R P}} \Big]+
	{\cal O}(\alpha'^2)~=~0,\label{a.2}
\end{eqnarray}
\vspace{-6mm}
\begin{eqnarray}
{\beta_{_{MN}}^B}&=& {{\beta_{_{MN}}^B}}^{\hspace{-2mm}(1)} +\alpha' ~ {{\beta_{_{MN}}^B}}^{\hspace{-2mm}(2)}+{\cal O}(\alpha'^2)~~~~~~~~ \nonumber\\
	&=&{\nabla}^{^P} H_{_{PMN}} -2  ({\nabla}^{^P}\Phi)  H_{_{MNP}}
	+\alpha' \Big[{\nabla}^{^P} H^{^{RS}}_{_{~~[M}}{\cal R}_{_{_{{N]} P RS}}} -\frac{1}{4}({\nabla}_{_P} H_{_{R MN}}) {H^2}^{^{PR}}\nonumber\\
	&&~~~~~~~~~~~~~+\frac{1}{2} ({\nabla}^{^P}H^{^{R Q}}_{_{~~[M}})H_{_{_{{N]} Q S }}} H_{_{P R}}^{^{~\;S}}\Big]
	+{\cal O}(\alpha'^2)~=~0,~~~~~~~\label{a.3}
\end{eqnarray}
\vspace{-6mm}
\begin{eqnarray}
{\beta^\Phi}&=& {{\beta^\Phi}}^{(1)} +\alpha' ~ {{\beta^\Phi}}^{(2)}+{\cal O}(\alpha'^2)\nonumber\\
	&=&2 \Lambda +2 {\nabla}^2 \Phi -4 ({\nabla} \Phi)^2+\frac{1}{6} H^{{2}}
	-\alpha' \Big[\frac{1}{4} {\cal R}_{_{MNRS}} {\cal R}^{^{MNRS}}
	-\frac{1}{12} ({\nabla}_{_M} H_{_{NRS}})
	({\nabla}^{^M} H^{^{NRS}})
	\nonumber\\
	&&~~~~~~~~~~~~~~-\frac{1}{8} H^{^{MN}}_{_{~~P}} H^{^{RS P}} {\cal R}_{_{MN RS}}-\frac{1}{4}{\cal R}_{_{MN}} {H^2}^{^{MN}} +\frac{3}{32} H^2_{_{MN}} {H^2}^{^{MN}}\nonumber\\
	&&~~~~~~~~~~~~~~+\frac{5}{96} H_{_{MN P }} H^{^M}_{_{~~RS}} H^{^{N R}}_{_{~~Q}} H^{^{P S Q}}\Big] +{\cal O}(\alpha'^2)=0,\label{a.4}
\end{eqnarray}
where ${\beta}^{(i)}$'s stand for the $i$-loop beta function equations. Moreover,
the covariant derivatives ${\nabla}_{_M}$, Ricci tensor ${\cal R}_{_{MN}}$ and Riemann tensor field
${\cal R}_{_{MNPQ}}$ are calculated from the metric $G_{_{MN}}$ that is also used for lowering and raising
indices, and $H_{_{MNP}}$ is the field strength corresponding to the $B$-field which is defined by
\begin{eqnarray}\label{a.1.1}
H_{_{MNP}} = \partial_{_{M}} B_{_{NP}} +\partial_{_{N}} B_{_{P M}}+\partial_{_{P}} B_{_{MN}}.
\end{eqnarray}
We have moreover introduced the conventional notations $H^2_{_{MN}}=H_{_{MPQ}}  H^{^{PQ}}_{_{~~N}}$, $H^2=H_{_{MNP}}
H^{^{MNP}}$,
${H^2}^{^{MN}} = H^{^{MPQ}} H_{_{PQ}}^{^{~~N}}$ and $({\nabla} \Phi)^2 =\partial_{_{M}} \Phi ~\partial^{^{M}} \Phi$.
Also, in equation \eqref{a.4},
$\Lambda$ is a cosmological constant.

In the following we shall investigate the conformal invariance conditions of the $\sigma$-models with metrics $AdS_2 $,
$AdS_{2}\times S^1$, $ AdS_3 $ and the metric of the analytic continuation of $AdS_{2}$ up to the second order in $\alpha'$
and also BTZ black hole up to the first order. Then, we will
study the non-Abelian T-duality of these metrics (except for BTZ) using the procedure mentioned in subsection \ref{Sec.II.1}.
As mentioned in Introduction section,  the Abelian T-duality of $AdS_2$ metric has been discussed in Ref. \cite{Tseytlin1}.
Here we are going to investigate the non-Abelian T-duality of this metric by PL T-duality procedure.
Let us now start with $AdS_2$ metric.

%%%%%%%%%%%%%%%%%%%%%%%%%%%%%%%%%%%%%%%%%%%%%%%%%%%%%%%%%%%%%%%%%%%%%%%%%%%%%%%%%%%%%%%%%%%%%%%%%%%%%%%%%%%%%
		\section{Non-Abelian T-duality of the $AdS_2$ background}
\label{Sec.III}
The simplest space of $AdS$ family is $AdS_2$ manifold whose metric may be expressed as
\begin{eqnarray}
	ds^2=\frac{l^2}{x^2}(- dt^2+dx^2)\label{(A)dS2}.
\end{eqnarray}
One can easily deduce that the field equations \eqref{a.2}-\eqref{a.4} to zeroth order in $\alpha'$ don't possess a $AdS_2$ solution with metric
\eqref{(A)dS2}. Notice that in the case of two-dimensional backgrounds such as $AdS_{2}$, the field strength $H$ is always zero.
Now we consider the $\alpha'$ terms to the equations \eqref{a.2}-\eqref{a.4}, but
neglect the $B$-field. In this case, the $B$-field equation of motion \eqref{a.3} is fulfilled.
Considering a constant dilaton field the resulting equations are then reduced to
two polynomial constraints:
\begin{eqnarray}
 1+\alpha' (-\frac{1}{l^2})=0,\label{(A)dS2.1}\\
2 \Lambda +\alpha' (-\frac{1}{l^4})=0.\label{(A)dS2.1.1}
\end{eqnarray}
As it can be seen from the above equations, the solutions we obtain are not under control.
Namely, from equation \eqref{(A)dS2.1} one takes the AdS radius to be the string length or $ \alpha'=l^2 $. In this way,
we can cancel one-loop terms against two-loop terms in the beta function equations.
Technically one can do this but it is not physically admissible. Similarly to our work, in the case of the
G\"{o}del spacetimes in string theory for the full ${\cal O}(\alpha')$ action,
Barrow and D\c{a}browski \cite{Barrow} found a simple relation between the angular velocity of
the G\"{o}del universe, $\Omega$, and the inverse string tension of the form $\alpha' =1/\Omega^2$ in the absence of the $B$-field.
Any way, it seems that the $\alpha'$ expansion is uncontrollable, since all orders now contribute equally.
In order to show that the expansion in the higher orders is uncontrollable, we calculate three-loop beta functions for the $AdS_2$ metric.
By following \cite{Jack}, in the absence of the field strength, the three-loop beta functions for a general theory are given by
\begin{eqnarray}\label{(A)dS2.2}
	{{\beta_{_{MN}}^G}}^{\hspace{-2.5mm}(3)}&=&\frac{1}{8} {\nabla}_{_P} {\cal R}_{_{M QRS}} {\nabla}^{^P}  {\cal R}_{_N}^{^{~QRS}}-\frac{1}{16}
{\nabla}_{_M} {\cal R}_{_{PQRS}} {\nabla}_{_N} {\cal R}^{^{PQRS}} -\frac{1}{2} {\cal R}_{_{MPQR}} {\cal R}_{_{NST}}^{^{~~~~R}} {\cal R}^{^{PTQS}}
\nonumber\\
	&&-\frac{3}{8}  {\cal R}_{_{MPQN}}  {\cal R}^{^{PRST}}   {\cal R}^{^{Q}}_{_{~RST}}
+\frac{1}{32}  {\nabla}_{_M}  {\nabla}_{_N} ({\cal R}_{_{PQRS}} {\cal R}^{^{PQRS}}),
\end{eqnarray}
\vspace{-3mm}
\begin{eqnarray}\label{(A)dS2.3}
{\beta^\Phi}^{(3)}&=&-\frac{3}{16} {\cal R}^{^{MPQR}} {\cal R}^{^{N}}_{_{~PQR}} {\nabla}_{_M}  {\nabla}_{_N} \Phi +\frac{1}{32}
{\cal R}_{_{MNPQ}} {\cal R}^{^{PQRS}} {\cal R}_{_{RS}}^{^{~~~MN}}~~~~~~~~~~~~~~~~~~~~~~~\nonumber\\
&&-\frac{1}{24} {\cal R}_{_{MNPQ}} {\cal R}^{^{RQSN}} {{\cal R}_{_{R}}^{^{~~P}}}
_{_{S}}^{~~M} +\frac{1}{64}  {\nabla}^{^M} \Phi {\nabla}_{_M} ({\cal R}_{_{PQRS}} {\cal R}^{^{PQRS}}).
\end{eqnarray}
Now we add  the three-loop beta functions \eqref{(A)dS2.2} and  \eqref{(A)dS2.3} to the field equations \eqref{a.2}-\eqref{a.4}.
Using the fact that the field strength is zero, the beta function equations to second order in $\alpha'$
for the $AdS_2$ metric with a constant dilaton reduce to the following polynomials,
\begin{eqnarray}
{\beta_{_{tt}}^G}=- {\beta_{_{xx}}^G} :=\frac{1}{x^2}\Big[1+\alpha' (-\frac{1}{l^2}) +\alpha'^2 (\frac{5}{4 l^4})\Big] + {\cal O}(\alpha'^3) =0,~\label{(A)dS2.4}\\ {\beta^\Phi}:=2 \Lambda +\alpha' (-\frac{1}{l^4}) +\alpha'^2 (-\frac{1}{4 l^6}) + {\cal O}(\alpha'^3) =0.~~~~\label{(A)dS2.1.5}
\end{eqnarray}
The above results render the $\alpha'$ expansion is uncontrollable,
and thus one can't guarantee the conformal invariance
of the $AdS_2$ metric.

To continue, in order to study the non-Abelian T-duality of the $AdS_2$ background one may obtain
Lie algebra generated by Killing vectors of \eqref{(A)dS2}.
The corresponding Killing vectors $ {k_{a}}$ of $AdS_2$ can be derived by solving Killing equations, $ \mathcal{L}_{_{k_{a}}}G_{_{MN}}=0$.
They are then read off
\begin{eqnarray}\label{kAdS_2}
	k_{1}&=&\frac{t^2+ x^2}{2}\partial_{t}
	+tx\partial_{x}\nonumber\\
	k_{2}&=&t\partial_{t}
	+x\partial_{x},\\
	k_{3}&=&- \partial_{t},\nonumber
\end{eqnarray}
such that the vectors $k_{1}$ and $k_{3}$ become everywhere timelike  except for  $x=0$. The Killing vectors \eqref{kAdS_2} also satisfy in the
 $sl(2 , \mathbb{R})$ algebra ($\cong$ $VIII$ Bianchi Lie algebra) with the following commutation relations
\begin{eqnarray}
	\left [ k_{2} ,k_{1}\right]= k_{1},~~~
	\left [ k_{1} ,k_{3}\right]= k_{2},~~~
	\left [ k_{2} ,k_{3}\right]= -k_{3}.\label{BVIII}
\end{eqnarray}
Note that this Lie algebra has only a two-dimensional non-Abelian subalgebra, $\mathcal{A}_2$, which can be defined by bases $ T_1=k_2 $ and $ T_2=k_1 $ (or $ T_1=-k_2 $ and $ T_2=k_3 $). As we will show below the Lie group corresponding to $\mathcal{A}_2$ acts freely and transitively on the $AdS_2 $ manifold.
In the next subsection we shall study the non-Abelian T-duality of the $AdS_2$ background
by constructing the PL T-dual $\sigma$-models stating from the Lie bialgebra $ (\mathcal{A}_2,2\mathcal{A}_1)$.
%%%%%%%%%%%%%%%%%%%%%%%%%%  3.1 A_2 %%%%%%%%%%%%%%%%%%%%%%%%%%%%%%%%%%%%%%%%%%%%%%%%%%%%%%%%%%%%%%%%%%%%%%

\subsection{Non-Abelian T-duality as PL T-duality on the semi-Abelian double $ (\mathcal{A}_2,2\mathcal{A}_1)$ }
\label{Sec.III1}
As mentioned above, the two-dimensional Lie subalgebra of \eqref{BVIII} is $\mathcal{A}_2$ with the following commutation relation:
\begin{eqnarray}
	\left [ T_{1} ,T_{2}\right]=T_{2},&\label{A_2}
\end{eqnarray}
where we have considered $ T_{1}=k_{2},T_{2}=k_{1} $.
In order to construct dualizable backgrounds we need to first investigate whether the action of the Lie group $\pmb{A_2}$ on $AdS_2$ manifold is free and transitive. Before proceeding further, let us have some review on free and transitive actions \cite{Nakahara}:\\\\
{\it Free action:}~
The free action of the Lie group $\pmb{G}$ on a manifold $\cal M$ is given by
\begin{eqnarray}
 g\circ x^M=x^M \Rightarrow g=e,\label{3.5}
\end{eqnarray}
for any $g=e^{\alpha^a T_a} \in  \pmb{G}$ and $x^M \in {\cal M}$ in which $T_a$'s are the bases of Lie algebra corresponding to $\pmb{G}$.
Considering the infinitesimal form of this action one easily finds that $ \xi \circ x^M=0 $ in which $ \xi=\alpha^a T_a $.
In order to have the free action one must expand the bases $T_a$ in terms of the Killing vectors of the manifold metric.
Then using
$ \xi \circ x^M=0 $ it should be concluded that $ \alpha^a=0 $, that is, $g=e$.\\\\
{\it Transitive action:}~
The transitive action of the Lie group $\pmb{G}$ on the manifold $\cal M$ means that for every two points $ x^M $ and $ {x'}^M $ of $\cal M$, there is a non-trivial element $ g $ of $\pmb{G}$ such that $ g\circ x^M={x'}^M $\footnote{Notice that if the action of the Lie group $\pmb{G}$ on the manifold $\cal M$ is free and also transitive, then the dimensions of both $\bf G$ and $\cal M$ are equal \cite{Nakahara}.}.
In order to turn the above definition into a computational method for determining the transitive action of $\pmb{G}$ on $\cal M$
one may expand the bases of Lie algebra $ \cal G $ of $\pmb{G}$ in terms of the Killing vectors $ k_a $ of $\cal M$, i.e, $ T_a=B_a^{~b}~ k_b$. Then it follows that
\begin{eqnarray}
T_a=A_a^{~M}\partial_M. \label{3.6}
\end{eqnarray}
Notice that from the linear independence of the bases $ T_a $ one concludes that the matrix $ A_a^{~M} $ must be invertible. Let $ g=e^{\alpha^aT_a} $ be an element of $\pmb{G}$. By considering infinitesimal form of $ g $ and by using $ g\circ x^M={x'}^M $ and also the invertibility condition on $ A_a^{~M} $ we then find that $ \alpha^a\ne 0 $. That is, $ g $ is a non-trivial element of $\pmb{G}$.
Thus, in order to have the transitive action of $\pmb{G}$ on $\cal M$, the matrix $ A_a^{~M} $ must be invertible.

In what follows we consider the action of Lie group $\pmb{A_2}$ with generators $ T_1=k_2, T_2=k_1 $ on $AdS_2$.
Considering $x^M=(x,t)$ we find $\xi=\big[\alpha^1 t +\frac{1}{2} \alpha^2  (t^2+ x^2)\big] \partial_{t} + (\alpha^1 x +\alpha^2 t x)\partial_{x}$.
Then, it follows from $ \xi \circ x^M=0 $ that $ \alpha^1=\alpha^2=0 $. Accordingly, one says that the action of $\pmb{A_2}$ on $AdS_2$ is free.
On the other hand,  inserting $ T_1=k_2=t\partial_{t}
+x\partial_{x} $ and $ T_2=k_1=\frac{1}{2}(t^2+ x^2)\partial_{t}+tx\partial_{x} $ into \eqref{3.6} we get
\begin{eqnarray}
	A_{a}^{~M}=\left(
	\begin{array}{cc}
		x & t \\
		tx & \frac{t^2+ x^2}{2} \\
	\end{array}
	\right),
\end{eqnarray}
such that $ det A\ne0 $. Therefore, the action of $\pmb{A_2}$ on $AdS_2$ is also transitive.
Now we are ready to find the non-Abelian target space dual of $AdS_2$ background.
Consider the Drinfeld double $(\mathcal{A}_2 , 2\mathcal{A}_1)$ \cite{Helvaty} (see, also, \cite{Harmonic}), where $ 2\mathcal{A}_1 $ is the two-dimensional Abelian Lie algebra. Making use of \eqref{DC} and \eqref{A_2} we obtain the corresponding non-zero Lie brackets
\begin{eqnarray}
\left [ T_{1} ,T_2\right]=T_2,~~~[ T_{1} ,\tilde{T}^{2}]=-\tilde{T}^{2},~~~ [ T_{2} ,\tilde{T}^{2}]=\tilde{T}^{1},\label{MC1}
\end{eqnarray}
where $(\tilde{T}^{1} , \tilde{T}^{2})$ are the bases of the dual Lie algebra $2\mathcal{A}_1$.
Choosing a convenient element of  $\pmb{A_2}$ as $g=e^{x_1T_1}e^{x_2T_2}$ and then using
\eqref{RIOF} we get the components of the right invariant one-forms
\begin{eqnarray}
	{R_{_M}}^{~a}=\left(
	\begin{array}{cc}
		1 & 0 \\
		0 & e^{x_1} \\
	\end{array}
	\right),
\end{eqnarray}
where $({x_1} , {x_2})$ stand for the coordinates of $\pmb{A_2}$.
It can be inferred from \eqref{abPi} and \eqref{MC1} that $ \varPi=0 $.
Thus by setting $ E_{0}(e) $ as
\begin{eqnarray}
	E_{0}(e)=l^2\left(
	\begin{array}{cc}
		1 & 0 \\
		0 & - 1 \\
	\end{array}
	\right),\label{E0}
\end{eqnarray}
the $ \sigma $-model \eqref{Sigma1} can be written in the following form
\begin{eqnarray}\label{AdS_2S}
	S=\frac{l^2}{2}\int d\sigma^+d\sigma^-\left[\partial_+x_1 \partial_-x_1- e^{2x_1}\partial_+x_2 \partial_-x_2\right].\label{SigmaA_2}
\end{eqnarray}
Comparing \eqref{AdS_2S} and the general form of $ \sigma $-model \eqref{Sigma2} we obtain that $ B_{_{MN}}=0 $ and
\begin{eqnarray}
	G_{_{MN}}=l^2 \left(
	\begin{array}{cc}
		1 &  0 \\
		0 & - e^{2x_1}  \\
	\end{array}
	\right).
\end{eqnarray}
Indeed the above metric can be transform to the $AdS_2$ metric \eqref{(A)dS2} by an appropriate coordinate transformation.
To find relation between coordinates of $AdS_2$, $(x,t)$, and group coordinates, $(x_1,x_2)$, one  must find the left invariant vector fields of $\pmb{A_2}$, giving us
\begin{eqnarray}
	V_1=\partial_{x_1}-x_2\partial_{x_2},&
	V_2=\partial_{x_2},
\end{eqnarray}
and then consider the transformation between these vector fields and $ \partial_M $ of $AdS_2$
\begin{eqnarray}
\left(
\begin{array}{c}
	V_1 \\
	V_2 \\
\end{array}
\right)=\left(
\begin{array}{cc}
	x & t \\
	tx & \frac{t^2 + x^2}{2} \\
\end{array}
\right) \left(
\begin{array}{c}
	\partial_x \\
	\partial_t \\
\end{array}
\right).
\end{eqnarray}
Finally, one can find the following transformation
\begin{eqnarray}
	x=\frac{2e^{x_1}}{x_2^2e^{2x_1}- 1},~~~~
	t=\frac{- 2x_2e^{2x_1}}{x_2^2e^{2x_1}- 1}.\label{Tr2}
\end{eqnarray}
To construct the dual $ \sigma $-model of \eqref{SigmaA_2}, in other words,  the dual space to $AdS_2$
we choose an element of $\pmb{2A_1}$ as  $\tilde{g}=e^{\tilde x_1\tilde T^1}e^{\tilde x_2\tilde T^2}$, where $ (\tilde x_1,\tilde x_2) $ are the coordinates of $\pmb{2A_1}$. Then by using relation \eqref{abPi} for the dual group and by applying \eqref{MC1} one can obtain the Poisson structure on $\pmb{2A_1}$, expressing
\begin{eqnarray}
	\tilde{\varPi}(\tilde{g})=\left(
	\begin{array}{cc}
		0 & -\tilde x_2  \\
		\tilde x_2 & 0 \\
	\end{array}
	\right).\label{Pit2}
\end{eqnarray}
Noting the fact that the dual Lie algebra is Abelian we find that $\tilde R_{_{M a}}=\delta_{_{M a}}$. Then
using \eqref{Et}, \eqref{E0},  \eqref{Pit2} together with \eqref{Sigmat1} the dual $ \sigma $-model corresponding to \eqref{AdS_2S} is worked out
\begin{eqnarray}
	\tilde S=\frac{1}{2}\int d\sigma^+d\sigma^-\frac{1}{l^4-\tilde{x}_2^2 }\left[l^2\partial_+\tilde x_1\partial_-\tilde x_1- l^2\partial_+\tilde x_2\partial_-\tilde x_2\right. \nonumber\\
	\left.- \tilde x_2\partial_+\tilde x_1\partial_-\tilde x_2+ \tilde x_2\partial_+\tilde x_2\partial_-\tilde x_1\right].\label{StAdS2}
\end{eqnarray}
One may compare the dual model \eqref{StAdS2} with dual version of  general $ \sigma $-model \eqref{Sigma2} to obtain
\begin{eqnarray}
	d\tilde s^2&=&\frac{l^2}{l^4-\tilde{x}_2^2 }\Big[d\tilde x_1^2- d\tilde x_2^2\Big],\label{3.18}\\
	\tilde B&=&- \frac{\tilde x_2}{l^4-\tilde{x}_2^2}~d\tilde x_1\wedge d\tilde x_2.\label{3.19}
\end{eqnarray}	
It is worthwhile to mention some specific and important properties of this dual solution.
The metric \eqref{3.18} is ill defined at the regions $ \tilde x_2=\pm l^2 $.
We can test whether  there are true singularities
by calculating the scalar curvature.
To be more specific, one may use the coordinate transformation
\begin{eqnarray}
\tilde x_1=l^2t,~~~~~~~~~~~~~\tilde x_2=\pm l^2\cosh{r},\label{3.20}
\end{eqnarray}
to write \eqref{3.18} and \eqref{3.19} in the following form
\begin{eqnarray}
	d\tilde s^2&=&l^2\Big[dr^2-\frac{1}{\sinh^2(r)}  dt^2\Big],\label{3.21}\\
	\tilde B&=&l^2\coth{r}~dt\wedge dr.\label{3.22}
\end{eqnarray}
The scalar curvature of the metric is
\begin{eqnarray}
	\tilde {\cal R} = -\frac{\big[3+\cosh (2 r )\big]}{l^2 \sinh^2(r)}.\label{3.23}
\end{eqnarray}
Thus, $ r=0 $ is a true singularity; moreover, one can show that this
singularity also appears in the Kretschmann scalar, which is, $\tilde {\mathcal{K}}={\tilde{\cal R}}^2$.
On the other hand, only the Killing vector of \eqref{3.21} is $-\partial_{t} $ whose norm is $ -l^2/\sinh^2(r)$,
hence, after the dualization only a timelike isometry is preserved.
Note that in the case of the dual metric \eqref{3.18} the $\alpha'$ expansion is also uncontrollable.
Namely, the dual background can't be conformally invariant.
\\\\
%%%%%%%%%%%%%%%%%%%%%%%%%%%%%%%%%%%%%%%%%%%%%%%%%%%%%%%%%%%%
%%%%%%%%%%%%%%%%%%%%%%%%%%%%%%%%%%%%%%%%%%%%%%%%%%%%%%%%%%%%
{\bf A note on the non-Abelian T-duality of the metric of the analytic continuation of $AdS_{2}$.}
The metric of the analytic continuation of $AdS_{2}$ in Poincar\'{e} coordinates can be derived by doing the Wick rotation $t \rightarrow i t$
on the metric \eqref{(A)dS2}, giving us
\begin{eqnarray}
	ds^2=\frac{l^2}{x^2}(dt^2+dx^2)\label{3.24.1}.
\end{eqnarray}
The Killing vectors of this metric are
\begin{eqnarray}\label{3.24.2}
	k_{1}&=&\frac{t^2- x^2}{2}\partial_{t}
	+tx\partial_{x}\nonumber\\
	k_{2}&=&t\partial_{t}
	+x\partial_{x},\\
	k_{3}&=&\partial_{t}.\nonumber
\end{eqnarray}
Unlike $ AdS_2 $, both Killing vectors $k_{1}$ and $k_{3}$ of \eqref{3.24.1} are spacelike.
One can easily check that the Lie algebra spanned by these vectors is $ IX $ Bianchi Lie algebra. Analogously,
only two-dimensional non-Abelian subalgebra of $ IX $ Bianchi is $\mathcal{A}_2$ which is defined by bases $ T_1=k_2 $ and $ T_2=k_1 $.
In this case,  the $\mathcal{A}_2$ Lie group also acts freely and transitively on the manifold defined by the metric \eqref{3.24.1}.
To find the non-Abelian target space dual of \eqref{3.24.1} we first construct the original $\sigma$-model on the semi-Abelian double $(\mathcal{A}_2 , 2\mathcal{A}_1)$  such that the metric of model can be turned into \eqref{3.24.1}. To this end, one must choose the constant matrix $E_{0}(e)$ as $E_{0}(e) =l^2 \mathbb{I} $,
where $\mathbb{I} $ is the $2 \times 2$ identity matrix, to obtain the metric of the model in the form of
\begin{eqnarray}\label{3.24.3}
	ds^2=l^2 \Big[{dx_1}^2 + e^{2x_1}{dx_2}^2\Big].
\end{eqnarray}
Under the coordinate transformation
\begin{eqnarray}\label{3.24.4}
	x=\frac{2e^{x_1}}{x_2^2e^{2x_1}+ 1},~~~~
	t=\frac{- 2x_2e^{2x_1}}{x_2^2e^{2x_1}+ 1},
\end{eqnarray}
the metric \eqref{3.24.3} can be turned into \eqref{3.24.1}.
The corresponding dual model can be constructed out by means of the procedure applied for $ AdS_2 $.
Finally, the dual background for the metric \eqref{3.24.1} reads
\begin{eqnarray}
	d\tilde s^2&=&\frac{l^2}{l^4+\tilde{x}_2^2 }(d\tilde x_1^2+ d\tilde x_2^2),\label{3.24.5}\\
	\tilde B&=&\frac{\tilde x_2}{l^4+\tilde{x}_2^2}~d\tilde x_1\wedge d\tilde x_2.\label{3.24.6}
\end{eqnarray}
Unlike the dual metric of $ AdS_2 $, there is no singularity for the dual metric of \eqref{3.24.1} as expected, because
these two examples are different in nature as Abelian T-duality performed in \cite{Tseytlin1} for both $ AdS_2 $ and $S^2$. There is only a spacelike Killing vector $\partial_{\tilde x_1} $ for the metric \eqref{3.24.5}.
So, after dualization only a spacelike isometry is preserved.
We also note that due to the uncontrollable $\alpha'$ expansion,
the conformal invariance of both original and dual backgrounds fail here as in $ AdS_2 $ case.

%%%%%%%%%%%%%%%%%%%%%%%%%%%%%%         Ads_2+S^1               %%%%%%%%%%%%%%%%%%%%%%%%%%%
\section{Non-Abelian T-duality of the $ AdS_2\times S^1 $ background}
\label{Sec.IV}
The $AdS_2\times S^1$ metric can be, in the coordinates $ (x, z, t) $, written as
\begin{eqnarray}
	ds^2=l^2dx^2+\frac{l^2}{z^2}(dz^2-dt^2).\label{AdS2S1}
\end{eqnarray}
Looking at the equations \eqref{a.2}-\eqref{a.4} together with \eqref{(A)dS2.2} and \eqref{(A)dS2.3},
one can check the conformal invariance conditions of the metric \eqref{AdS2S1}.
Hence, the vanishing of the beta function equations up to three-loop order for the $ AdS_2\times S^1 $ metric with a zero $B$-field and dilaton field $ \Phi=c_1x+c_2 $, where $ c_i $'s are some constant parameters,
reduce to the two polynomials,
\begin{eqnarray}
{\beta_{_{tt}}^G}=- {\beta_{_{zz}}^G} :=\frac{1}{z^2}\Big[1+\alpha' (-\frac{1}{l^2}) +\alpha'^2 (\frac{5}{4 l^4})\Big] + {\cal O}(\alpha'^3)
=0,~\label{(A)dS2.4}\\
{\beta^\Phi}:=2 \Lambda -\frac{4 {c_1}^2}{l}+\alpha' (-\frac{1}{l^4}) +\alpha'^2 (-\frac{1}{4 l^6}) + {\cal O}(\alpha'^3) =0.~~~~\label{(A)dS2.1.5}
\end{eqnarray}
From the above equations one concludes that $\alpha'$ expansion
is uncontrollable as in the $AdS_2$ metric. Therefore the $AdS_2\times S^1$ background fails to satisfy the beta function equations
which indicates that the corresponding $\sigma$-model is not Weyl invariant, i.e. does not define a critical string theory
in the usual sense.

In the following, in order to investigate the non-Abelian T-duality of the $AdS_2\times S^1$ metric we need to
obtain the Lie algebra generated by the Killing vectors of \eqref{AdS2S1}.
The metric \eqref{AdS2S1} admits the following Killing vectors
\begin{eqnarray}
	k_{1}&=&\partial_{x},\nonumber\\
	k_{2}&=&tz\partial_{z}+(\frac{t^2+z^2}{2})\partial_{t},\\
	k_{3}&=&z\partial_{z}+t\partial_{t},\nonumber\\
	k_{4}&=&-\partial_{t}.\nonumber
	\label{kAdS2S1}
\end{eqnarray}
The Lie algebra generated by these Killing vectors is
\begin{eqnarray}\label{4Dim}
	\left [ k_{2} ,k_{3}\right]=-k_{2},&
	\left [ k_{2} ,k_{4}\right]=k_{3},&
	\left [ k_{3} ,k_{4}\right]=-k_{4}.
\end{eqnarray}
According to the norms of the Killing vectors $ |k_1|^2=l^2, |k_2|^2=-l^2 \left(t^2-z^2\right)^2/({4 z^2})$ and  $|k_4|^2=-l^2/z^2$ we find that
the vector $ k_1 $ is everywhere spacelike while $ k_2 $ and $ k_4 $ are everywhere
timelike except for  $z=0$.
Indeed, the Lie algebra given by \eqref{4Dim} is isomorphic to the $gl(2 , \mathbb{R})$.
One can show that there are two classes of three-dimensional subalgebras of the isometry algebra of the $AdS_2\times S^1$ metric isomorphic to
 the Bianchi Lie algebras $ III$ and $ VIII$ which are denoted by
$ III_{.i}$ and $ VIII_{.i} $, respectively.

\begin{table}[h]
	\caption{ \scriptsize {Free and transitive actions of three-dimensional isometry Lie subgroups on the $AdS_2\times S^1$ }}
	\centering
	\begin{tabular}{| p{7em} | p{15em} | p{4em} | p{5em}|}
		\hline
		
		\scriptsize  {Bianchi Type}  & \scriptsize {Subalgebra} & \scriptsize {Free}&\scriptsize{Transitive}\\
		
		\hline
		
		\scriptsize $ {III_{.i}} $&\scriptsize $ Span \{ k_3,k_2,k_1\} $& \scriptsize Yes & \scriptsize Yes \\
		
		\hline
		
		\scriptsize $ {VIII_{.i} }$&\scriptsize $ Span \{ -k_3,k_4,2k_2\} $& \scriptsize Yes & \scriptsize No \\
		
		\hline
	\end{tabular}
	\label{table:1}
\end{table}

%%%%%%%%%%%%%%%%%%%%%%%%%%%%%%%%%%%%%%%%%%%%%%%%%%%%%%%%%%%%%%%%%%%%%
On the other hand, one easily shows that the
Lie group corresponding to $ III_{.i}$, $\pmb{III_{.i}}  $, acts freely and transitively on $AdS_2\times S^1$ space, while for the Lie group $\pmb{VIII_{.i}} $ we have a free action only. The results are summarized in Table \ref{table:1}.

%%%%%%%%%%%%%%%%%%%%%%%%%%%%%%%%%%%%%%%%%%%%%%%%%%%%%%%%%%%%%%%%%%%%%%%%%%%%%%%

\begin{table}[h!]
		\caption{\scriptsize {The non-Abelian T-duality results of the $ AdS_2\times S^1 $ background}}
		\centering
		\begin{tabular}{| p{4.2em} | p{7.79em} | p{6.43em} | p{9.20em} |  }
			\hline
			
			\multirow{2}{4.2em}{\scriptsize  {~~~Double} }
			& {\scriptsize ~~~~~~~~~Non-zero}
			& {\scriptsize {~~Relation of $ T_a $'s  }}
			&\multirow{2}{9em} {\scriptsize  {$ ~~~~~~~~~~~~~~A_a^{~M} $} }
			\\
			
			&{\scriptsize Commutation relations}
			&{\scriptsize {to Killing vectors}}
			&
			\\
			%%%%%%%%%%%%%%%%%%%%%%%%%%%%%%%%%%%%%%%%%%%
			\hline
			
			\multirow{3}{4.2em}{\scriptsize $ (III_{.i},3\mathcal{A}_1)$}
			&\scriptsize $ [T_1,T_2]=T_2, $
			&\scriptsize $ T_1=k_3, $
			& \multirow{3}{8em}{\scalebox{0.6}{~~~~~~$
				\left(
					\begin{array}{ccc}
						0 & t & z \\
						0 & \frac{1}{2} (t^2+z^2) & tz \\
						1 & 0 & 0 \\
					\end{array}
					\right)\nonumber
					$}}
			\\

			&\scriptsize $ [T_1,\tilde T^2]=-\tilde T^2, $
			&\scriptsize $ T_2=k_2, $
			&
			\\

			&\scriptsize $ [T_2,\tilde T^2]=\tilde T^1. $
			&\scriptsize $ T_3=k_1. $
			&
			\\
			\hline
		\end{tabular}			
			%%%%%%%%%%%%%%%%%%%%%%%%%%%%%%%%%%%%%%%%%%%%%%%%%%%%%%%%%%	
				%%%%%%%%%%%%%%%%%%%%%%%%%%%%%%%%%%%%%%%%%%%%%%%%%%%%%%%%%%	
				\begin{tabular}{| p{4.2em} | p{6.40em} | p{6.90em} | p{10.10em} |  }
\hline
\multirow{2}{4.2em} {\scriptsize $ ~~~~E_0(e) $}
& \scriptsize ~~~~~Coordinate
& \multirow{2}{7.00em}{\scriptsize Original background}
&\multirow{2}{7.00em}{\scriptsize ~~Dual background}
\\

&\scriptsize ~~~transformation
&
&
\\
\hline

%%%%%%%%%%%%%%%%%%%%%%%%%%%%%%%%%%%%%%%%%%%%%
\multirow{3}{4.2em}{\scalebox{0.6}{$
		l^2\left(
		\begin{array}{ccc}
			1 & 0 & 0 \\
			0 & -1 & 0 \\
			0 & 0 & 1 \\
		\end{array}
		\right)\nonumber
		$}}
&\scriptsize $ x=x_3, $
&\scriptsize $ ds^2=l^2\left(dx_1^2+dx_3^2\right. $
& \scriptsize $ d\tilde s^2=\frac{d\tilde x_3^{2}}{l^2}+\frac{l^2}{l^{4}-\tilde x_2^{2}}(d\tilde x_1^{2}-d\tilde x_2^{2}),  $
\\

&\scriptsize $ z=\frac{2 {\mathrm e}^{x_1}}{{\mathrm e}^{2 x_1} x_2^{2}-1}, $
&\scriptsize $ \left.-e^{2x_1}dx_2^2\right), $
&\scriptsize $ \tilde B=\frac{\tilde x_2}{l^{4}-\tilde x_2^{2}}d\tilde x_2\wedge d\tilde x_1. $
\\

&\scriptsize $ t=-\frac{2 {\mathrm e}^{2 x_1} x_2}{{\mathrm e}^{2 x_1} x_2^{2}-1}. $
&\scriptsize $ B=0. $
&
\\
\hline
		\end{tabular}
		
		\label{table:2}
\end{table}

%%%%%%%%%%%%%%%%%%%%%%%%%%%%%%%%%%%%%%%%%%%%%%%%%%%%%%%%%%%%%%%%%%%%%%%%%%%%%%%
Similar to the T-dual $\sigma$-models construction for the  $AdS_2$  metric, which was represented in preceding section,
we find the non-Abelian target space dual including the metric and $B$-field of the $ AdS_2\times S^1 $ background.
As shown in Table \ref{table:2}, we have constructed T-dual $\sigma$-models\footnote{Note that we can also obtain the original background from the PL T-duality in the presence of spectator fields. To this end, one may use two-dimensional Lie group $ \pmb{A_2} $ with the coordinates $ (x_1,x_2) $ and choose a spectator with the coordinate $ x_3 $. So far, our findings show that the Lie group $ \pmb{A_2} $ is wealthy. Recently, in order to study the non-Abelian T-duality of
the G\"{o}del spacetimes \cite{Godel} we have constructed the T-dual $\sigma$-models on
the manifold ${\cal M} \approx O \times \pmb{A_2}$ where $\pmb{A_2}$ acts freely on ${\cal M}$ while $O$ is the two-dimensional orbit of $\bf G$ in ${\cal M}$
(see, also, \cite{{EMR13},{Eghbali.2}}).} based on the semi-Abelian
double $ (III_{.i},3\mathcal{A}_1)$ \cite{{Snobl},{RHR}} by a convenient choice
of the constant matrix $E_0(e)$. It has been shown that there is an appropriate coordinate transformation that turns original $\sigma$-model
into the $ AdS_2\times S^1 $ background. As it is seen from the dual solution,
we don't expect to see any dramatic change in physical properties of $ AdS_2\times S^1 $ versus $AdS_2$ alone.

%%%%%%%%%%%%%%%%%%%%%%%%%%%%%%%%%%%%%%%%%%%%%%%%%%%%%%%%%%%%%%%%%%%%%%%%%%%%%%%%%

%%%%%%%%%%%%%%%%%%%%%%%%%%%%%%%%%%%%%%%%%%%%%%%%%%%%%%%%%%%%%%%%%%%%%%%%%%%%%%%%%%

\section{Non-Abelian T-duality of the $AdS_3$ background}
\label{Sec.V}
According to \eqref{1} the metric of $AdS_3$ in coordinates $ (t, x, y) $ is given by\footnote{It should be remarked that the
non-Abelian T-dualization of $AdS_3$ background has been already discussed in Ref. \cite{maleki}.
There, the $AdS_3$ metric has been applied in  Poincar\'{e} coordinates as in \eqref{2}, hence,
the forms of their Killing vectors and Lie algebra spanned by them are different from ours (Eqs. \eqref{kAdS_3} and \eqref{6Dim}). Although subalgebras of the isometry subgroups of those are isomorphic to our results, the form of the resulting dual backgrounds is different from ours (Table 4).
Most importantly, we have, here, investigated
the spacetime structure of T-dual findings for $ AdS_3 $ background and also some their physical interpretations
by introducing some convenient coordinate transformations, while
these cases are not seen in \cite{maleki}.
}
\begin{eqnarray}
	ds^2=\frac{l^2}{y^2}(-dt^2+dx^2+dy^2).\label{AdS3}
\end{eqnarray}
For the metric \eqref{AdS3} with zero B-field and a constant dilaton field, the vanishing of the three-loop beta function equations
are reduced to two polynomial constraints:
\begin{eqnarray}
{\beta_{_{xx}}^G}= {\beta_{_{yy}}^G}=-{\beta_{_{tt}}^G} :=-\frac{2}{y^2}
\Big[1+\alpha' (-\frac{1}{l^2}) +\alpha'^2 (\frac{3}{l^4})\Big] + {\cal O}(\alpha'^3)
=0,~\label{(A)dS2.4}\\
{\beta^\Phi}:=2 \Lambda + \alpha' (-\frac{3}{l^4}) +\alpha'^2 (-\frac{1}{2 l^6})
+ {\cal O}(\alpha'^3) =0.~~~~\label{(A)dS2.1.5}
\end{eqnarray}
In this case, the $\alpha'$ expansion can't be also controlled in the same way as in the previous cases.
Nevertheless, as mentioned in \cite{Horowitz2} the metric \eqref{AdS3} in the presence of the $B$-field $ B_{tx}=-l^2/y^2  $ with a constant dilaton field
satisfy the one-loop beta function equations
provided that $\Lambda ={2}/{l^2}$.
It can be also useful to comment on the fact that one can verify the field equations \eqref{a.2}-\eqref{a.4} up to two-loop order for the aforementioned solution provided that $\Lambda = (2 l^2 +4\alpha')/l^4$.
It's worth mentioning $AdS_3$ inherits the isometries of the embedding space that preserve the hyperboloid. The
group of rotations+boosts in a $4D$ geometry with signature $(+, +, -, -)$ is $\bf {SO(2, 2)}$, so
we expect this to be the isometry group of $AdS_3$\footnote{In general the $AdS_d$
isometry group is $\bf{SO(d-1, 2)}$  \cite{Tseytlin1}.}. In this section we will confirm this.

We are interested in metrics that admit
at least three independent Killing vectors because they can be interpreted as
T-dualizable backgrounds for $\sigma$-models in three dimensions.
In order to investigate the non-Abelian T-duality of the $AdS_3$ we need
all three-dimensional subalgebras of Killing vectors that generate group of isometries acting freely and transitively on the $AdS_{3}$ manifold.
Metric \eqref{AdS3} has a number of symmetries important for the construction of
the dualizable $\sigma$-models. It admits the following Killing vectors
\begin{eqnarray}
	k_{1}&=&\frac{t^2+x^2-y^2}{2}\partial_{x}+tx\partial_{t}+xy\partial_{y},\nonumber\\
	k_{2}&=&\frac{t^2+x^2+y^2}{2}\partial_{t}+ty\partial_{y}+tx\partial_{x},\nonumber\\
	k_{3}&=&x\partial_{x}+t\partial_{t}+y\partial_{y},\nonumber\\
	k_{4}&=&t\partial_{x}+x\partial_{t},\nonumber\\
	k_{5}&=&\partial_{x},\nonumber\\
	k_{6}&=&-\partial_{t}.~~~~~\label{kAdS_3}
\end{eqnarray}
It is easily shown that the vectors $k_{1}$ and $k_{5}$ are everywhere spacelike except for  $y=0$, while
$k_{2}$ and $k_{6}$ stay everywhere timelike except for  $y=0$.
The Lie algebra spanned by Killing vectors \eqref{kAdS_3} is isomorphic to the $ so(2,2) $ Lie algebra
with nonzero commutation relations
\begin{eqnarray}\label{6Dim}
	\left [ k_{3} ,k_{1}\right]=\left [ k_{4} ,k_{2}\right]=k_{1},~~~
	\left [ k_{4} ,k_{1}\right]=\left [ k_{3} ,k_{2}\right]=k_{2},~~~\nonumber\\
	\left [ k_{5} ,k_{1}\right]=\left [ k_{2} ,k_{6}\right]=k_{3},~~~
	\left [ k_{1} ,k_{6}\right]=\left [ k_{5} ,k_{2}\right]=k_{4},~~~\\
	\left [ k_{5} ,k_{3}\right]=\left [ k_{4} ,k_{6}\right]=k_{5},~~~
	\left [ k_{6} ,k_{3}\right]=\left [ k_{4} ,k_{5}\right]=k_{6}.~~~\nonumber
\end{eqnarray}
One can check that the three-dimensional Bianchi Lie algebras $ III_{.i}, V, VI_0, VI_q,$ and $ VIII_{.i} $ are Lie subalgebras of \eqref{6Dim} such that all the corresponding Lie subgroups (except for the $ \pmb{VI_0} $ Lie group) act freely and transitively on the $AdS_3$ space. The results are summarized in Table \ref{table:3}.\footnote{To obtain the commutation relations of the semi-Abelian doubles generated by the Bianchi Lie algebras  \cite{{Snobl},{RHR}} of Table \ref{table:3}, one must use \eqref{DC} and \eqref{6Dim} together with the basis represented in terms of the linear combination of Killing vectors.}
 \begin{table}[h]
 	\caption{ \scriptsize {Free and transitive actions of three-dimensional isometry Lie subgroups on $AdS_3$. }}
 	\centering
 	\begin{tabular}{| p{5em} | p{19em} | p{3em} | p{3em}|}
 		\hline
 		
 		\scriptsize  {Bianchi Type}  & \scriptsize  {~~~~~~~~~~~~~~Subalgebra} & \scriptsize {Free}&\scriptsize {Transitive}\\
 		
 		\hline
 		
 		\scriptsize $ {III_{.i}} $&\scriptsize $ Span \{ T_1=-k_3,T_2=k_5+k_6,T_3=k_3+k_4\} $& \scriptsize Yes & \scriptsize Yes \\
 		
 		\hline
 		
 		\scriptsize $ V $&\scriptsize $ Span \{T_1= k_4,T_2=2(k_1+k_2),T_3=k_5+k_6\} $& \scriptsize Yes & \scriptsize Yes \\
 		
 		\hline
 		
 		\scriptsize $ {VI_0} $&\scriptsize $ Span \{ T_1=-k_6,T_2=k_5,T_3=k_4\} $& \scriptsize Yes & \scriptsize No \\
 		
 		\hline
 		
 		\scriptsize $ {VI_q} $&\scriptsize $ Span \{ T_1=k_4-qk_3,T_2=k_6,T_3=-k_5\} $& \scriptsize Yes & \scriptsize Yes \\
 		
 		\hline
 		
 		\scriptsize $ {VIII_{.i}} $&\scriptsize $ Span \{T_1= (k_3+k_4)/2,T_2=(k_1+k_2)/2,T_3=k_6-k_5\} $& \scriptsize Yes & \scriptsize Yes \\
 		
 		\hline
 	\end{tabular}
 	\label{table:3}
 \end{table} \\

Similar to the construction of T-dual $\sigma$-models for $AdS_2 $ and $ AdS_2\times S^1 $ backgrounds which were represented in preceding sections and
also by applying the results of Table \ref{table:3} we find the non-Abelian target space duals of $ AdS_3 $.
For the sake of clarity the results obtained in this section are summarized in Table \ref{table:4};
we display the metrics and $B$-fields corresponding to both original and dual backgrounds,
together with the transformation between $ AdS_3 $ coordinates and group ones, as well as the form of constant matrix $E_0(e)$.

%%%%%%%%%%%%%%%%%%%%%%%%%%%%%%%%%%%%%%%%%%%%%%%%%%%%%%%%%
\begin{table}[h!]
	\begin{threeparttable}
		\caption{ The non-Abelian T-duality results of the $ AdS_3 $ background}
		\centering
		\begin{tabular}{| p{4.2em} | p{5em} | p{7.95em} | p{10.38em} | p{12.99em} | }
			\hline
			
			\multirow{2}{4.2em}{\scriptsize  {~~~Double} }
			&\multirow{2}{5em} {\scriptsize $~~~~~~ E_0(e) $}
			& {\scriptsize {~~~~~~~Coordinate }}
			&\multirow{2}{9em} {\scriptsize  {~~~~~~Original Backgrounds} }
			&\multirow{2}{16em}{\scriptsize {~~~~~~~~~~~~~Dual Backgrounds}}
			\\
			
			&
			&{\scriptsize {~~~~~transformation }}
			&
			&
			\\
			%%%%%%%%%%%%%%%%%%%%%%%%%%%%%%%%%%%%%%%%%%%
			\hline
			
			\multirow{3}{4.2em}{\scriptsize $ (III_{.i},3\mathcal{A}_1)$}
			& \multirow{3}{8em}{\scalebox{0.6}{$
					l^2\left(
					\begin{array}{ccc}
						0 & -1 & 0 \\
						-1 & 0 & 2 \\
						0 & 2 & 1 \\
					\end{array}
					\right)\nonumber
					$}}
			&\scriptsize$ t=\frac{e^{-x_1}(e^{2x_3}+1)-2x_2}{2},$
			&{\scriptsize $ ds^2=l^2 \left[\right.4 e^{x_1} dx_2 dx_3 +{dx_3}^2$}
			&\scriptsize $ d\tilde s^2=\frac{2l^2}{l^4-\tilde{x}_2^2}(2d\tilde x_1^2-d\tilde x_1d\tilde x_2+2d\tilde x_1d\tilde x_3)$
			\\

			&
			&\scriptsize$ x=\frac{e^{-x_1}(e^{2x_3}-1)+2x_2}{2},$
			&\scriptsize$  -2 e^{x_1} dx_1 dx_2 \left.\right], $
			& \scriptsize $+l^{-2}d\tilde x_3^2,$
			\\

			&
			&\scriptsize$ y=e^{x_3-x_1}.$
			&\scriptsize $ B=0. $
			&\scriptsize $ \tilde B=\frac{2\tilde x_2}{ l^4-\tilde{x}_2^2}d\tilde x_1\wedge d\tilde x_3. $
			\\

			%%%%%%%%%%%%%%%%%%%%%%%%%%%%%%%%%%%%%%%%%%%%%%%%%%%%%%%%%%	
			\hline
			
			\multirow{3}{4.2em}	{\scriptsize $ (V,3\mathcal{A}_1)$}
			& \multirow{3}{8em}{\scalebox{0.6}{$
					l^2\left(
					\begin{array}{ccc}
						1 & 0 & 0 \\
						0 & 0 & \frac{1}{2} \\
						0 & \frac{1}{2} & 0 \\
					\end{array}
					\right)
					$}}
			&\scriptsize$ t=\frac{4 e^{-2 x_1}+1}{-4 x_2}-x_3, $
			&{\scriptsize $ ds^2=l^2 \left( e^{2 x_1} dx_2 dx_3+dx_1^2\right),$}
			&\scriptsize $ d\tilde s^2=\frac{1}{\Gamma_1} \Big[l^2d\tilde x_1^2   + 4 l^2 d\tilde x_2d\tilde x_3   $
			\\

			&
			&\scriptsize$ x=\frac{4e^{-2 x_1}-1}{4x_2}+x_3, $
			&{\scriptsize $ B=0.$}
			&\scriptsize$  -\frac{4}{l^2}(\tilde x_2d\tilde x_3 -\tilde x_3d\tilde x_2)^2\Big], $
			\\
			
			&
			&\scriptsize$ y=-\frac{e^{-x_1}}{x_2}. $
			&
			&\scriptsize $ \tilde B=\frac{2}{\Gamma_1}(\tilde x_3d\tilde x_1\wedge d\tilde x_2+\tilde x_2d\tilde x_1\wedge d\tilde x_3). $
			\\
			%%%%%%%%%%%%%%%%%%%%%%%%%%%%%%%%%%%%%%%%%%%%%%%%%%%%%%%%%%	
			\hline

			&\multirow{6}{8em}	{\scalebox{0.6}{$
					l^2\left(
					\begin{array}{ccc}
						q^2 & 0 & 0 \\
						0 & -1 & 0 \\
						0 & 0 & 1 \\
					\end{array}
					\right)
					$}}
			&
			&
			&\scriptsize $  d\tilde s^2=\frac{1}{\Gamma_2}\Big[l^2d\tilde x_1^2$
			\\

			&
			&\scriptsize$ t=-x_2, $
			&\scriptsize $ ds^2=l^2 \left(q^2 dx_1^2\right.  $
			&\scriptsize$ -\big(q^2l^2+\frac{ \left(q\tilde x_3-\tilde x_2\right)^2 }{l^2}\big)d\tilde x_2^2$
			\\
			
			{\scriptsize $ (VI_q,3\mathcal{A}_1)$}
			&
			&\scriptsize$ x=-x_3, $
			&\scriptsize $ \left. +e^{2 q x_1} \left(dx_3^2-dx_2^2\right)\right), $
			&\scriptsize$ +\big(q^2l^2-\frac{ \left(q\tilde x_2-\tilde x_3\right)^2 }{l^2}\big)d\tilde x_3^2 $
			\\
			
			{\scriptsize $q\ne0,\pm 1$}
			&
			&\scriptsize$ y=e^{- q x_1}. $
			&\scriptsize $ B=0. $
			&\scriptsize$ +2\frac{\left(q\tilde x_3-\tilde x_2\right) \left(q\tilde x_2-\tilde x_3\right)}{l^2}d\tilde x_2d\tilde x_3\Big] ,  $
			\\
			
			&
			&
			&
			&\scriptsize $ \tilde B=\frac{1}{\Gamma_2}\Big[(\tilde x_3-q \tilde x_2)d\tilde x_1\wedge d\tilde x_2$
			\\

			&
			&
			&
			&\scriptsize$  -(\tilde x_2-q \tilde x_3)d\tilde x_1\wedge d\tilde x_3\Big].  $
			\\
			
			%%%%%%%%%%%%%%%%%%%%%%%%%%%%%%%%%%%%%%%%%%%%%%%%%%%%%%%%%%	
			\hline
			
			\multirow{5}{4.2em}{\scriptsize $ (VIII_{.i},3\mathcal{A}_1)$}
			&\multirow{5}{8em}	{\scalebox{0.6}{$
					\frac{l^2}{4}\left(
					\begin{array}{ccc}
						1 & 0 & 0 \\
						0 & 0 & 2 \\
						0 & 2 & 0 \\
					\end{array}
					\right)
					$}}
			&
			&
			&\scriptsize $ d\tilde s^2=\frac{4}{\Gamma_3}\left[ \left(l^4-16 \tilde x_1^2\right)d\tilde x_1^2\right.$
			\\

			&
			&\scriptsize$ t=\frac{-4 x_2 x_3-4 e^{x_1}-1}{4 x_2}, $
			&\scriptsize $ ds^2=\frac{l^2}{4}  \left(4 e^{-x_1} dx_2 dx_3 \right. $
			&\scriptsize$ \left.-16   (\tilde x_3d\tilde x_2 +\tilde x_2d\tilde x_3 )\right.\tilde x_1d\tilde x_1 -4 \tilde x_3^2d\tilde x_2^2 $
			\\

			&
			&\scriptsize$ x=\frac{-4 x_2 x_3-4 e^{x_1}+1}{4 x_2}, $
			&\scriptsize $ \left.+dx_1^2\right), $
			&\scriptsize$ \left.+  \left(l^4-8 \tilde x_2 \tilde x_3\right)d\tilde x_2 d\tilde x_3-4\tilde x_2^2 d\tilde x_3^2 \right], $
			\\
			
			&
			&\scriptsize$ y=-\frac{e^{\frac{x_1}{2}}}{x_2}. $
			&\scriptsize $ B=0. $
			&\scriptsize$\tilde B=\frac{8l^2}{\Gamma_3}\Big[-\tilde x_3d\tilde x_1\wedge d\tilde x_2 $
			\\
			
			&
			&
			&
			&\scriptsize$ +\tilde x_2d\tilde x_1\wedge d\tilde x_3-\tilde x_1d\tilde x_2\wedge d\tilde x_3\Big].  $
			\\
			
			\hline
		\end{tabular}
		\begin{tablenotes}
			\scriptsize
			{\item Note: $ \Gamma_1=l^4+4 \tilde x_2 \tilde x_3 $, $ \Gamma_2=q^2l^4+(1-q^2) \left(\tilde x_2^2-\tilde x_3^2\right) $ and $ \Gamma_3= l^2 \left[l^4- 16  (\tilde x_1^2 + \tilde x_2 \tilde x_3)\right]$ }
		\end{tablenotes}
		\label{table:4}
	\end{threeparttable}
\end{table}
%%%%%%%%%%%%%%%%%%%%%%%%%%%%%%%%%%%%%%%%%%%%%%%%%%%%%%%%%%%%%%%%%%%%
In order to better understand of the spacetime structure of T-dual findings for $ AdS_3 $ background and also some their physical interpretations
we use some coordinate transformations that make the metrics simpler.
Below we discuss the dual backgrounds for each case separately.\\\\
$\bullet$ {\it The dual background on the double $ (III_{.i},3\mathcal{A}_1)$}:
As shown in Table \ref{table:4}, the dual metric obtained by this double has apparent singularities at the regions ${\tilde x}_2 = \pm l^2$.
Indeed, these are the coordinate singularities in the metric. To remove them one may use the change of coordinates
\begin{eqnarray}
\tilde x_1=l^2y,~~~~~~\tilde x_2=-l^2\tanh{z},~~~~~~~~~\tilde x_3=l^2x,\label{5.6}
\end{eqnarray}
to rewrite the dual background in the form
\begin{eqnarray}
	d\tilde s^2&=&l^2\Big[dx^2+4\cosh^2{z} (dy^2+ dy dx)+2dydz\Big],\label{5.7}\\
	\tilde B&=&l^2 \sinh(2z) ~ dx \wedge dy.\label{5.8}
\end{eqnarray}
The scalar curvature and Kretschmann scalar corresponding to the metric are, respectively, given by
\begin{eqnarray}
	\tilde {\cal R}&=&-\frac{1}{l^2} \big[1+7 \cosh (4 z)\big],\label{5.9}\\
	\tilde {\mathcal{K}}&=& \frac{1}{l^4} \Big[26 \cosh (4 z)+\frac{19}{2} (\cosh (8 z)+3)\Big].\label{5.10}
\end{eqnarray}
As it is seen from equations \eqref{5.7}, \eqref{5.9} and \eqref{5.10}, the singular points are not true points.
In addition, note that the metric \eqref{5.7} also possesses two independent Killing vectors
$\partial_x$ and $\partial_y - \partial_x$
which have the norm $ l^2 $, thus, they are everywhere spacelike.
Therefore, the duality has not here involved the timelike directions.
In the case of the conformal invariance conditions of background given by
\eqref{5.7} and \eqref{5.8} we have checked that this background does not satisfy the field equations \eqref{a.2}-\eqref{a.4} up to two-loop order.\\\\
$\bullet$ {\it The dual background on the double $ (V , 3\mathcal{A}_1)$}:
In this case, it is simply shown that the field strength corresponding to the $B$-field represented in Table \ref{table:4} is zero.
Hence, if we introduce the coordinate transformation
\begin{eqnarray}
\tilde x_1=l^2x,~~~~~~\tilde x_2=\frac{l^2}{2} y e^t,~~~~~~\tilde x_3=\frac{l^2}{2} y e^{-t},\label{5.11}
\end{eqnarray}
then, the dual background turns into
\begin{eqnarray}
	d\tilde s^2&=&l^2\left[\frac{ dx^2+dy^2}{y ^2+1}- y ^2 dt^2\right],\label{5.12}\\
	\tilde B&=&0. \label{5.13}
\end{eqnarray}
Here we have ignored the total derivative terms that appeared in the $B$-field part.
As it is seen from \eqref{5.12}, the singularities appeared in dual metric have been removed by coordinate transformation \eqref{5.11}.
Solving the field equations \eqref{a.2}-\eqref{a.4}
for the metric \eqref{5.12} with zero field strength
one concludes that there is no suitable dilaton field
to satisfy these equations.
It is also interesting to see that the metric \eqref{5.12} possesses two independent Killing vectors
$-\partial_t$ and $\partial_x$ which are timelike and spacelike, respectively. Thus, in this case, the duality involves the timelike directions.\\\\
$\bullet$ {\it The dual background on the double $ (VI_q,3\mathcal{A}_1)$}:
In this case one may use the coordinate transformation
\begin{eqnarray}
\tilde x_1=l^2t,~~~~~~~~\tilde x_2=2 l^2 e^{\frac{u}{2}} \cosh(\frac{v}{2}),~~~~~~~~\tilde x_3=-2 l^2 e^{\frac{u}{2}} \sinh(\frac{v}{2}),\label{5.14}
\end{eqnarray}
then, the dual background related to this double yields
\begin{eqnarray}
	d\tilde s^2&=& \frac{l^2  }{\bigtriangleup}\Big[-dt^2 + e^{u}  (q^2 +4 e^{u}) du^2 +
q^2 e^{u} (4 e^{u}-1) dv^2 -8q e^{{2u}} du dv\Big],\label{5.15}\\
	\tilde B&=&\frac{2l^2 e^{u} }{\bigtriangleup}\Big[q dt\wedge du - dt\wedge dv\Big],\label{5.16}
\end{eqnarray}
where $\bigtriangleup=4 \left(q^2-1\right) e^{u}-q^2 $. For the metric \eqref{5.15} one finds that the scalar curvature is
\begin{eqnarray}
\tilde {\cal R}= \frac{-32 (q^2-1)^2 e^{2u} -16 (q^2-1) e^{u}  +2 q^2(q^2-5)}{l^2 \bigtriangleup^2}.\label{5.17}
\end{eqnarray}
Therefore, here the singularity is true and represents itself as the curve  $u=-\ln \left(4 [1-q^{-2}]\right) $.\\
\\
$\bullet$ {\it The dual background on the double $ (VIII_{.i},3\mathcal{A}_1)$}: In this case, the from of the metric represented in Table \ref{table:4} is a bit complicated.
In order to get the simpler form of the metric one may consider the following coordinate transformation
\begin{eqnarray}
\tilde x_1=\frac{l^2}{4}e^{z},~~~~~~\tilde x_2=\frac{l^2}{4} e^{y+\sqrt{2}x},~~~~~~\tilde x_3=\frac{l^2}{4} e^{y-\sqrt{2}x}.\label{5.18}
\end{eqnarray}
Under the above transformation, the background becomes
\begin{eqnarray}
	d\tilde s^2&=&\frac{l^2}{2 (e^{2 y }+e^{2 z }-1)} \Big[e^{2 y } dx^2+ e^{2 (y +z )}dy dz+e^{3 y } \sinh (y ) dy^2+e^{3 z }  \sinh (z )dz^2\Big],\label{5.19}\\
	\tilde B&=&\frac{l^2  e^{2 y +z }}{ 2 \sqrt{2} \big(e^{2 y }+e^{2 z }-1)}(dz\wedge dx-dy\wedge dx\big).\label{5.20}
\end{eqnarray}
By calculating the scalar curvature corresponding to the metric \eqref{5.19} one concludes
that the singularity appeared in the metric, which represents itself as the curve  $ e^{2 y}+e^{2 z}=1 $, is true.
In this case, the metric just has one Killing vector which is $ 2\partial x $ with the norm $ 2 l^2 e^{2 y }/{(e^{2 y }+e^{2 z }-1)} $.
Accordingly, the behavior of Killing vector changes between spacelike and timelike regions as we pass through the singularity curve.

%%%%%%%%%%%%%%%%%%%%%%%%%%%%%%%%%%%%%%%%%%%%%%%%%%%%%%%%%%%%%%%%%%%%%%%%%%%%%
%%%%%%%%%%%%%%%%%%%%%%%%%%%%%%%%%%%%%%%%%%%%%%%%%%%%%%%%%%%%%%%%%%%%%%%%%%%%%
%%%%%%%%%%%%%%%%%%%%%%%%%%%%%%%%%%%%%%%%%%%%%%%%%%%%%%%%%%%%%%%%%%%%%%%%%%%%%
\section{Abelian T-duality of the BTZ black hole up to two-loop order}
\label{Sec.VI}
As announced in Introduction section, the BTZ black hole metric \cite{Banados} with the following $B$-field and dilaton
can be considered as a solution for the equations of motion of the low energy string effective action
\cite{Horowitz2}
\begin{eqnarray}\label{6.1}
	ds^2&=&(M-\frac{r^2}{l^2})dt^2-Jdtd\varphi+r^2d\varphi^2+\Big(\frac{J^2}{4r^2} +\frac{r^2}{l^2}-M\Big)^{-1}dr^2,\nonumber\\
	B&=&\frac{r^2}{l}~d\varphi\wedge dt,\nonumber\\
	\Phi&=&b,
\end{eqnarray}
for some constant $b$.
By studying Buscher-duality of this solution \cite{Horowitz2}, it was shown that the BTZ black hole solution is, under the Abelian T-duality, equivalent to
the charged black string solution discussed in \cite{Horowitz1}. Then, in \cite{EMR13} by investigating
the non-Abelian T-duality of the BTZ vacuum solution
it was shown that the non-Abelian T-duality transformation
relates the BTZ vacuum solution with no horizon and no curvature singularity  to
a solution with a single horizon and a curvature singularity.
In this section, we study the Abelian T-duality of BTZ background up to $\alpha'$-corrections
when the dualizing is implemented by the shift of directions $\varphi$ and $t$.
To this end, we use the T-duality rules at two-loop order derived by KM \cite{KM}.
Before proceeding further, let us review the $\alpha'$-corrected T-duality
rules of KM in the next subsection.

\subsection{A review of Abelian T-duality up to $\alpha'$-corrections}
\label{Abelian}

The two-loop $\sigma$-model corrections to the
T-duality map in string theory by using the effective action approach were obtained by KM \cite{KM} \footnote{Notice that one can also derive
the Abelian T-duality rules of KM from the $\alpha'$-corrected rules
of non-abelian T-duality proposed by Borsato and  Wulff \cite{Wulff22}.}.
They had found the explicit form for the ${\cal O } {(\alpha')}$ modifications
of the lowest order duality transformations by focusing on backgrounds that
have a single Abelian isometry.
Following Ref. \cite{KM}, here we consider the reduced metric $g_{{\mu\nu}}$, antisymmetric field $b_{{\mu\nu}}$
and dilaton $\Phi$ of the $d$-dimensional spacetime as
\begin{eqnarray}
	d s^2&=&G_{_{MN}} dX^{^{M}} dX^{^{N}} = g_{{\mu\nu}} dx^{\mu} dx^{\nu} + e^{2 \sigma}~ (d\underline{x} +V_{\mu} dx^{\mu})^2,\label{6.2}\\
	B &=& \frac{1}{2} B_{_{MN}} dX^{^{M}} \wedge dX^{^{N}} = \frac{1}{2} b_{\mu\nu} dx^{\mu} \wedge dx^{\nu}
	+\frac{1}{2} W_{\mu} V_{\nu}  dx^{\mu} \wedge dx^{\nu} + W_{\mu}   dx^{\mu} \wedge d\underline{x},\label{6.3}\\
	\Phi &=& \hat{\phi}  +\frac{1}{2} \sigma,\label{6.4}
\end{eqnarray}
where $V_{\mu}$ is the standard Kaluza-Klein
gauge field coupled to the momentum modes of the theory, and $W_{\mu}$ is the other gauge field
coupling to the winding modes. It has been assumed that the isometry direction we
want to dualize is implemented by a shift of a coordinate $ \underline{x} $.
Furthermore, ${\sigma}$ and $\hat{\phi} $ in equation \eqref{6.4} are some scalars fields.
The relations to identify the fields of the dimensional reduction are given by
\begin{eqnarray}
	V_{{\mu}}&=& \frac{G_{{\mu \underline{x}}} }{G_{{\underline{x} \underline{x}}}},
	~~~~~~~~~~~~~~~~~~~~~~~~~W_{{\mu}} = B_{{\mu \underline{x}}},\label{6.5}\\
	\sigma&=& \frac{1}{2} \log G_{{\underline{x} \underline{x}}},~~~~~~~~~~~~~~~~~~~~~\hat{\phi} = \Phi  -\frac{1}{2} \sigma,\label{6.6}\\
	g_{{\mu\nu}}&=& G_{{\mu\nu}} - \frac{G_{{\mu \underline{x}}} G_{{\nu \underline{x}}}}{G_{{\underline{x}  \underline{x}}}},
	~~~~~~~~~~~b_{{\mu\nu}} = B_{{\mu\nu}} + \frac{G_{{\underline{x} [\mu}} B_{{\nu] \underline{x}}}}{G_{{\underline{x}  \underline{x}}}}.\label{6.6.1}
\end{eqnarray}
As first demonstrated in Ref. \cite{Wulff}, for applying rules of KM, one first needs to implement
the field redefinitions to go from HT scheme  to that of KM. The field redefinitions are given by \cite{Wulff}
\begin{eqnarray}
	G_{_{MN}}^{^{(HT)}}&=&G_{_{MN}}^{^{(KM)}} + \alpha'({\cal R}_{_{MN}}-\frac{1}{2}H^2_{_{MN}}),\label{6.7}\\
	B_{_{MN}}^{^{(HT)}}&=&B_{_{MN}}^{^{(KM)}} + \alpha'(-H_{_{MNP}}  \nabla^{^{P}}\Phi),\label{6.8}\\
	\Phi^{^{(HT)}} &=& \Phi^{^{(KM)}}  + \alpha'(-\frac{3}{32}H^2 + \frac{1}{8}{\cal R}-\frac{1}{2}(\nabla \Phi)^2).\label{6.9}
\end{eqnarray}
The equations of two-loop T-duality transformation in the KM scheme that we will use are \cite{KM}
\begin{eqnarray}
	{\tilde \sigma}&=&-\sigma + \alpha'[(\nabla \sigma)^2 +\frac{1}{8} (e^{2\sigma} Z+e^{-2\sigma} T)],\label{6.10}\\
	{\tilde V}_{\mu}&=&W_{\mu} + \alpha'[W_{{\mu \nu}} \nabla^{\nu}\sigma+  \frac{1}{4} h_{\mu\nu\rho} V^{\nu\rho} e^{2\sigma}],\label{6.11}\\
	{\tilde W}_{\mu}&=&V_{\mu} + \alpha'[V_{{\mu \nu}} \nabla^{\nu}\sigma-  \frac{1}{4} h_{\mu\nu\rho} W^{\nu\rho} e^{-2\sigma}],\label{6.12}\\
	{\tilde b}_{{\mu \nu}} &=& { b}_{{\mu \nu}}  + \alpha'\Big[V_{\rho[\mu} W^{\rho}_{~\nu]}+ (W_{[\mu\rho} \nabla^{\rho}\sigma +\frac{1}{4} e^{2\sigma} h_{[\mu\rho\lambda} V^{\rho \lambda}) V_{\nu]}~~~~~~\nonumber\\
	~~~~&&+(V_{[\mu\rho} \nabla^{\rho}\sigma -\frac{1}{4} e^{-2\sigma} h_{[\mu\rho\lambda} W^{\rho \lambda}) W_{\nu]}\Big],\label{6.13}
\end{eqnarray}
where
\begin{eqnarray}
	W_{{\mu \nu}}&=&\partial_{\mu} W_\nu -\partial_{\nu} W_\mu,\label{6.14}\\
	V_{{\mu \nu}}&=&\partial_{\mu} V_\nu -\partial_{\nu} V_\mu,\label{6.15}\\
	h_{\mu\nu\rho}&=& H_{\mu\nu\rho}-3 W_{[\mu\nu} V_{\rho]},\label{6.16}
\end{eqnarray}
also,
\begin{eqnarray}
	Z_{{\mu \nu}}&=&V_{{\mu \rho}} V_{\nu}^{~\rho},~~~~~~~~~~~~~~~~~~Z = Z_{\mu}^{~\mu},\label{6.17}\\
	T_{{\mu \nu}}&=&W_{{\mu \rho}} W_{\nu}^{~\rho},~~~~~~~~~~~~~~~~T = T_{\mu}^{~\mu}.\label{6.18}
\end{eqnarray}
All the lowering and raising of the indices  will be done with respect to the reduced metric $g_{{\mu \nu}}$. Note that after perform two-loop T-dulaity transformation one must return to HT scheme by use of \eqref{6.7}-\eqref{6.9}.
In the next subsections we will apply this method in order to study the Abelian T-duality of BTZ background up to two-loop order.
%%%%%%%%%%%%%%%%%%%%%%%%%%%%%%%%%%%%%%%%%%%%%%%%%%%%%%%%%%%%%%%%%%%%%%%%%%%%%%%%%%%%%%%%

%%%%%%%%%%%%%%%%%%%%%%%%%%%%%%%%%%%%%%%%%%%%%%%%%%%%%%%%%%%%%%%%%%%%%%%%%%%%%%%%%%%%%%%%%%%%%%%%%%%
\subsection{Dualizing with respect to the coordinate $ \varphi $}
\label{VI.2}
As we mentioned earlier, background \eqref{6.1} is a solution for the field equations
\eqref{a.2}-\eqref{a.4} up to zeroth order in $\alpha'$. In addition, one can show that
\eqref{6.1} satisfies \eqref{a.2}-\eqref{a.4} up to first order in $\alpha'$
if the following relation holds between the constants $l, \alpha'$  and $\Lambda$:
\begin{eqnarray}
\Lambda = \frac{2(2\alpha'+l^2)}{l^4}.
\end{eqnarray}
We note that solution \eqref{6.1} has been obtained in the HT scheme.
In order to use the field redefinitions
in equations \eqref{6.7}-\eqref{6.9} we need to write \eqref{6.1} in the KM scheme, giving us
\begin{eqnarray}
	{ds^2}^{^{(KM)}} &=&(M-\frac{r^2}{l^2}) d{t}^2 +\Big(\frac{J^2}{4 r^2}+\frac{r^2}{l^2}-M\Big)^{-1} dr^2-J ~ dt d\varphi +r^2 d\varphi^2\nonumber\\
	&&+\frac{2}{l^2}\Big[(\frac{r^2}{l^2}- M) dt^2 +{J} dt d\varphi-\big(\frac{J^2}{4 r^2}+\frac{r^2}{l^2}-M\big)^{-1} dr^2-r^2 d\varphi^2 \Big] \alpha',\nonumber\\
	H^{^{(KM)}}&=& \frac{2r}{l} dt \wedge dr \wedge  d\varphi,\nonumber\\
	\Phi^{^{(KM)}}  &=&b-\frac{3 }{2 l^2}\alpha'.\label{6.28}
\end{eqnarray}
In this way we are ready to obtain the shifting coordinates and then perform the KM T-duality transformations.
Here, the isometry we want to dualize is that the shift of the $\varphi$ coordinate, i.e., $\underline{x} =\varphi$.
It should be remarked that the rules of T-duality are derived assuming that the coordinate to be dualized is spacelike.
Fortunately, for the metric \eqref{6.1}, $G_{\underline{x x}} = G_{\varphi \varphi}$ is positive. In fact,  we are faced with a spacelike case.
Comparing \eqref{6.28} with equations \eqref{6.2}-\eqref{6.4} or using \eqref{6.5}-\eqref{6.6.1} one concludes that only non-zero components of
the fields $V_{_\mu}$ and $W_{_\mu}$ are
\begin{eqnarray}\label{6.22.1.1}
	V_t=-\frac{J}{2 r^2},~~~~~~W_t=-\frac{r^2}{l}.
\end{eqnarray}
Furthermore,
\begin{eqnarray}\label{6.23.1.1}
b_{\mu\nu}&=&0,\nonumber\\
{\sigma}^{^{(KM)}} &=&  \ln r-\frac{\alpha'}{l^2}+\mathcal{O}(\alpha'^2),\nonumber\\
\hat{\phi} &=& b-\frac{\ln r}{2}-\frac{\alpha'}{l^2}+\mathcal{O}(\alpha'^2).
\end{eqnarray}
Then, one may use \eqref{6.14}, \eqref{6.15} together with \eqref{6.22.1.1} to obtain only non-zero components of
$V_{_{\mu \nu}}$ and $W_{_{\mu \nu}}$ as follows
\begin{eqnarray}\label{6.24.1.1}
V_{rt}=\frac{J}{r^3},~~~~~~
W_{tr}=\frac{2r}{l}.
\end{eqnarray}
In this case one also gets that $h_{\mu\nu\rho} =0$.
Thus, by using \eqref{6.17}, \eqref{6.18} together with \eqref{6.24.1.1}, functions $Z$ and $T$ are, up to first order of $ \alpha' $, obtained to be
\begin{eqnarray}
	Z=-\frac{2 J^2}{r^6}-\frac{8J^2}{l^2r^6} \alpha' +\mathcal{O}(\alpha'^2),~~~~T=-\frac{8 r^2}{l^2}-\frac{32r^2}{l^4} \alpha' +\mathcal{O}(\alpha'^2).
\end{eqnarray}
Now, one may apply the above results to the two-loop T-duality transformation equations \eqref{6.10}-\eqref{6.13}
to get the dual solution in the KM scheme. Finally, by employing field redefinitions in equations \eqref{6.7}-\eqref{6.9} and
after some calculations we can write the dual solution in the HT scheme, giving
\begin{eqnarray}
	{{\tilde ds}^2}^{^{(HT)}} &=&(M-\frac{J^2}{4 r^2}) d{t}^2 + \Big(\frac{J^2}{4r^2} +\frac{r^2}{l^2}-M\Big)^{-1}dr^2-\frac{2}{l}~ dt d\underline{\tilde x} +\frac{1}{r^2} d{\underline{\tilde x}}^2-\alpha'\frac{2 }{r^2}dr^2,\nonumber\\
	{\tilde H}^{^{(HT)}}&=&-\frac{J}{r^3} ~dt \wedge dr \wedge d\underline{\tilde x},\label{6.25.2}\\
	{\tilde \Phi}^{^{(HT)}}&=& b-\ln r+\big(\frac{M}{2 r^2}-\frac{1}{l^2}\big)\alpha'.\nonumber
\end{eqnarray}
As expected this dual background satisfies the field equations \eqref{a.2}-\eqref{a.4} up to the first order in $\alpha'$
so that the relation between constants $\tilde \Lambda$, $l$ and $\alpha'$ may be expressed as $ \tilde \Lambda=(4 \alpha' + 2l^2)/l^4 $.
To be more explicit, choose two constants $r_+$, $r_-$ and introduce new coordinates
\begin{eqnarray}
t&=&\frac{l(\hat x-\hat t)}{(r_+^2-r_-^2)^{1/2}},\nonumber\\
\underline{\tilde x}&=&\frac{r_-^2\hat x-r_+^2\hat t}{(r_+^2-r_-^2)^{1/2}},\\
r^2&=&l\hat r\nonumber,
\end{eqnarray}
where the constants $M$ and $J$ are related to $r_{\pm}$ by
\begin{eqnarray}
M=\frac{(r_+^2+r_-^2)}{l^2},~~~~~~~~ J=\frac{2r_+r_-}{l}.
\end{eqnarray}
Then, background \eqref{6.25.2} becomes
\begin{eqnarray}
{{\tilde ds}^2}&=&-(1-\frac{\mathbb{M}}{\hat r})d\hat t^2+(1-\frac{\mathbb{Q}^2}{\mathbb{M}\hat r})d\hat x^2+(1-\frac{\mathbb{M}}{\hat r})^{-1}(1-\frac{\mathbb{Q}^2}{\mathbb{M}\hat r})^{-1}\frac{l^2d\hat r^2}{4\hat r^2}-\alpha'\frac{d\hat r^2 }{2\hat r^2},\nonumber\\
{\tilde H}&=&-\frac{\mathbb{Q}}{\hat r^2} ~d\hat t \wedge d\hat r \wedge d{\hat x},\nonumber\\
{\tilde \Phi}&=& b-\frac{1}{2}\ln (l\hat r)+\big(\frac{\mathbb{M}^2+\mathbb{Q}^2}{2 l^2\mathbb{M}\hat r}-\frac{1}{l^2}\big)\alpha',\label{BS}
\end{eqnarray}
where $ \mathbb{M}=r_+^2/l $ and $ \mathbb{Q}=J/2 $.
It should be noted that in the absence of $\alpha'$-corrections, solution \eqref{BS} is nothing but three-dimensional
charged black string solution  \cite{Horowitz1}.
Here we were able to obtain $\alpha'$-corrections sentences for the
charged black string solution using the $\alpha'$-corrected T-duality
rules derived by KM.
As it is seen from the metric,  its components are ill defined at $\hat r=0$, $\hat r = {\hat r}_{+} = \mathbb{M}$
and $\hat r = {\hat r}_{-} = {\mathbb{Q}^2}/{\mathbb{M}}$.
In order to determine true singularity one may first calculate the scalar curvature of the metric. It is, up to the first order in
$\alpha'$, read off
\begin{eqnarray}\label{6.37}
\hspace{-4mm}{\tilde {\cal R}}&=&\frac{2\big[(2\hat r-7 \mathbb{M})\mathbb{Q}^2 +2\hat r \mathbb{M}^2\big]}{\mathbb{M} l^2 \hat r^2}~\nonumber\\
&&\hspace{-10mm}+\frac{\Big[-12 {\hat r}^2 (\mathbb{M}^4 +\mathbb{Q}^4)-4 \mathbb{M}^2 \mathbb{Q}^2 (11 \mathbb{Q}^2 +13 \hat r^2)+
4\mathbb{M} \hat r (\mathbb{M}^2+\mathbb{Q}^2) (13 \mathbb{Q}^2 +2 \hat r^2)\Big]}{\mathbb{M}^2 l^4 \hat r^4} \alpha'
+\mathcal{O}(\alpha'^2).~~
\end{eqnarray}
Thus, $\hat r=0$ is a true curvature singularity; moreover, the horizons of the metric, ${\hat r}_{+}, {\hat r}_{-} $,
are at the same location as the charged black string without $\alpha'$-corrections.

%%%%%%%%%%%%%%%%%%%%%%%%%%%%%%%%%%%%%%%%%%%%%%%%%%%%%%%%%%%%%%%%%%%%%%%%%%%%%

\subsection{Dualizing with respect to the coordinate $t$}
\label{VI.3}
Here the isometry coordinate we want to dualize is the $t$ coordinate, i.e., $\underline{x} =t$.
For the metric \eqref{6.1}, the region $r < {M}^{1/2} l$ defines an ergosphere, in which the
asymptotic timelike Killing field $\partial/ {\partial t}$ becomes spacelike.
For this region, $G_{\underline{x x}} = G_{tt}$ is positive; consequently, we are faced with a spacelike case.
Comparing \eqref{6.28} with equations  \eqref{6.2}-\eqref{6.4} or using \eqref{6.5}-\eqref{6.6.1} we find that the only non-zero components of
$V_{_\mu}$ and $W_{_\mu}$ are
\begin{eqnarray}
	V_\varphi=\frac{J }{2 ( \frac{r^2}{l^2}-M )},~~~~~~~~~~~~~W_\varphi=\frac{r^2}{l},\label{6.29.1.1}
\end{eqnarray}
in addition, $ b_{\mu\nu}=0 $ and
\begin{eqnarray}
	{\sigma}^{^{(KM)}} =  \frac{1}{2} \ln\left(M-\frac{r^2}{l^2}\right)-\frac{\alpha'}{l^2}+\mathcal{O}(\alpha'^2),\nonumber\\
	\hat{\phi}   = b-\frac{1}{4} \ln \left(M-\frac{r^2}{l^2}\right)-\frac{\alpha'}{l^2}+\mathcal{O}(\alpha'^2).\label{t6.29}
\end{eqnarray}
Then, by using \eqref{6.14}-\eqref{6.16} together with \eqref{6.29.1.1} one gets that
\begin{eqnarray}
	V_{\varphi r}=\frac{Jl^2r}{(r^2-Ml^2)^2},~~~~~
	W_{r\varphi}=\frac{2r}{l},~~~~~~h_{\mu\nu\rho}=0.\label{6.32.1}
\end{eqnarray}
In order to obtain the functions $Z$ and $T$ one must use \eqref{6.32.1} together with equations
\eqref{6.14} and \eqref{6.15}.
Thus, these functions are, up to the first order of $ \alpha' $, read off
\begin{eqnarray}
Z&=&\frac{2 J^2l^2 }{\left(-l^2 M+r^2\right)^3}(1+\frac{4\alpha'}{l^2})+\mathcal{O}(\alpha'^2),\nonumber\\
T&=&\frac{8 \left(-l^2 M+r^2\right)}{l^4}(1+\frac{4\alpha'}{l^2})+\mathcal{O}(\alpha'^2).
\end{eqnarray}
Applying these results to the two-loop T-duality transformation equations \eqref{6.10}-\eqref{6.13}
one can get the dual solution in the KM scheme. Finally, by employing field redefinitions in equations \eqref{6.7}-\eqref{6.9}
we can write the dual solution in the HT scheme, giving us
\begin{eqnarray}
	{{\tilde ds}^2}^{^{(HT)}} &=&\Big(M-\frac{r^2}{l^2}\Big)^{-1} d\underline{\tilde x}^2 + \Big(\frac{J^2}{4r^2}-M+ \frac{r^2}{l^2}\Big)^{-1} dr^2+\frac{2l r^2}{l^2 M-r^2}~ d\varphi d\underline{\tilde x} \nonumber\\
&&-\frac{l^2 \left(J^2-4 M r^2\right)}{4 (l^2 M- r^2)} d{\varphi}^2-\alpha'\frac{2  r^2}{\left(r^2-l^2 M\right)^2}dr^2,\nonumber\\
	{\tilde H}^{^{(HT)}}&=&-\frac{J l^2 r}{\left(r^2-l^2 M\right)^2} ~d\underline{\tilde x} \wedge dr \wedge d\varphi,\nonumber\\
	{\tilde \Phi}^{^{(HT)}}&=& b-\frac{1}{2} \ln \big(M-\frac{r^2}{l^2}\big)-\frac{\left(l^2 M - 2 r^2\right)}{2 l^2 (l^2 M-r^2)}\alpha'.\label{t6.30}
\end{eqnarray}
One can check that the dual background \eqref{t6.30} is conformally
invariant up to the first order in $\alpha'$ provided that $ \tilde \Lambda=(4 \alpha' + 2l^2)/l^4 $.
To better understand this dual solution, we diagonalize the metric. Let
\begin{eqnarray}
	\underline{\tilde x}&=&   \frac{1}{\sqrt{2}} \Big[-({\cal J}+l M ) \phi+\frac{({\cal J}-l M)} {l} t\Big],\nonumber\\
	\varphi &=& \sqrt{2} ~(\phi + \frac{t}{l}),\nonumber\\
	r^2&=& {l\bar r},
\end{eqnarray}
where ${\cal J}=\sqrt{l^2 M^2-J^2}$. Then, the solution \eqref{t6.30} becomes
\begin{eqnarray}\label{6.36}
{{\bar {ds}}^2} &=&
\frac{{\cal J}}{l (\bar r-l M)} \Big[\big(-2 \bar r+l M -{\cal J}\big) dt^2 + \big(2 \bar r-l M -{\cal J}\big) l^2 d\phi^2\Big]  \nonumber\\
&&+\Big[\frac{l^2}{J^2+4 \bar r (\bar r-l M)}-\frac{\alpha'}{2 (\bar r-l M)^2}\Big] d\bar r^2,\nonumber\\
{\bar H}^{^{(HT)}}&=&-\frac{J {\cal J}}{\left(\bar r-l M\right)^2} ~dt \wedge d{\bar r} \wedge d\phi,\nonumber\\
	{\bar \Phi}^{^{(HT)}}&=& b-\frac{1}{2} \ln \big(M-\frac{\bar r}{l}\big)-\frac{\left(2 {\bar r}^2 -l M\right)}{2 l^2 ({\bar r}^2 -l M)}\alpha'.
\end{eqnarray}
The metric components are ill defined at $\bar r=l M$ and $\bar r = {\bar r}_{\pm} \equiv (l M \pm {\cal J})/2$.
Now, one can test whether
these are true singularities by looking at the scalar curvature, which up to the first order in $\alpha'$ is
\begin{eqnarray}\label{6.37}
{\bar {\cal R}}&=&\frac{8 l M (l M-\bar r)-7 J^2}{2 l^2 (\bar r-l M)^2}\nonumber\\
&&+\frac{\Big[-11 J^4+4 J^2 \left(6 l^2 M^2+l M \bar r-7 \bar r^2\right)-16 l M (\bar r-l M)^2 (l M+2 \bar r)\Big]}{4 l^4 (\bar r-l M)^4} \alpha'
+\mathcal{O}(\alpha'^2).
\end{eqnarray}
Thus, only $\bar r=l M$ is a true curvature singularity. The surface $\bar r = {\bar r}_{+}$  is the outer
horizon, while the $\bar r = {\bar r}_{-}$ is another surface inside the black string called the inner
horizon. According to definition ${\cal J}$, it is simply followed from ${\cal J} < l M$ that ${\bar r}_{-} <{\bar r}_{+} <l M$.
In fact, the true singularity lies outside the outer horizon here.
As mentioned in the Introduction, similar to our black string, in Ref.  \cite{suggett2} it has been shown that there is a singularity outside of a Schwarzschild black hole.
The metric given by \eqref{6.36} also possesses two independent Killing vectors $k_1 = \frac{1}{l{\cal J}} \partial_\varphi$ and
$k_2= -\frac{l}{{\cal J}} \partial_t$ with the norms
\begin{eqnarray}\label{6.38}
	|k_1|^2 = \frac{{\cal J}+l M-2 \bar r}{l{\cal J} (l M-\bar r)},~~~~~~~~~~
	|k_2|^2=\frac{l \left({\cal J}-l M+2 \bar r\right)}{{\cal J} (l M-\bar r)} .
\end{eqnarray}
The Killing vector $k_1$ becomes timelike for the ranges
${\bar r}_{+} <\bar r < l M$ which lies outside the outer horizon. It is also spacelike at infinity.
The Killing vector $k_2$ which is timelike at infinity becomes spacelike for ${\bar r}_{-} <{\bar r} <lM $
which includes the space between two horizons.

Let us investigate the asymptotic behavior of the metric.
Note that for large $\bar r$ it is not possible to similarly to the $t$ and $\phi$ coordinates fix the
overall scaling of the $\bar r$  as $\bar r$ goes to infinity, since the metric asymptotically approaches
${(l^2-2 \alpha') d \bar{r}^2}/{4 \bar{r}^2}$. Therefore, for large $r$ the black string solution \eqref{6.36} approaches the following
asymptotic solution
\begin{eqnarray}
{{\bar{ds}}^2} &=& -  d {\tau}^2 +   d {u}^2 + d \rho^2,\nonumber\\
{\bar \Phi} &=& {\bar b}-\frac{\rho}{l} -(\frac{\rho +l}{l^3}) \alpha',
~~~~~~~{{\bar H}} = 0, \label{6.40.1}
\end{eqnarray}
for some constant $\bar b$. Here we have set
\begin{eqnarray}
\bar{r}=e^{\frac{2\rho}{\sqrt{l^2-2 \alpha'}}} \simeq e^{\frac{2 \rho}{l}} \Big[1+ \frac{2 \rho}{l^3} \alpha' + \mathcal{O}(\alpha'^2)\Big],~~~~~~~
\phi = \frac{l}{\sqrt{2 l  {\cal J}}} u,
~~~~~~~~t = \sqrt{\frac{l}{2 {\cal J}}} ~\tau. \label{6.41.1}
\end{eqnarray}
Now, one verifies the field equations \eqref{a.2}-\eqref{a.4} up to two-loop order for
the asymptotic solution \eqref{6.40.1} with the same condition of the conformal invariance of background \eqref{t6.30}.
The above result shows that the Abelian T-duality transformation of KM changes
the asymptotic behavior of solutions from $AdS_3$ to flat.

%%%%%%%%%%%%%%%%%%%%%%%%%%%%%%%%%%%%%%%%%%%%%%%%%%%%%%%%%%%%%%%%%%%%%%%%%%%%%
%%%%%%%%%%%%%%%%%%%%%%%%%%%%%%%%%%%%%%%%%%%%%%%%%%%%%%%%%%%%%%%%%%%%%%%%%%%%%
\section{Conclusions}
\label{Sec.VII}
In order to study the non-Abelian T-duality of the metrics of Riemannian manifolds
one may use the isometry subgroups of the metrics.  Sufficient condition for that is that the dimension of the isometry subgroups of the metric is equal to the dimension of the Riemannian manifold and its action on the manifold is transitive and free. Using this fact we have shown that for the metrics of $AdS$ families such isometry subgroups exist and the metrics can be dualized by the PL T-duality transformation.
We have shown that the Lie subgroup ${\bf A_2}$ corresponding
to non-Abelian two-dimensional subalgebra $\mathcal{A}_2$
acts freely and transitively on $AdS_2 $ manifold. In this way, it has been found the dual of the
$AdS_2$ background stating from the Lie bialgebra $ (\mathcal{A}_2 , 2\mathcal{A}_1)$.
Our results show that the dual metric of the $AdS_2$ has a true singularity. In fact,
T-duality takes that singular region to regular region as was the case with the 2D black holes \cite{Dijgkraaf}.
As we have shown the Lie algebra generated by Killing vectors of $AdS_2 \times S^1$ is isomorphic to the $gl(2 , \mathbb{R})$. Then we have found that only the
Lie group corresponding to the $ III_{.i}$ Bianchi Lie algebra (as subalgebra of the $gl(2 , \mathbb{R})$) acts freely and transitively on $AdS_2\times S^1$ space. Accordingly, we have determined the metric and $B$-field dual to the $AdS_2 \times S^1$.
In the case of $AdS_3$, there are five classes of three-dimensional subalgebras of the isometry algebra of the $AdS_3$ metric isomorphic to
the Bianchi Lie algebras $ III, V, VI_0, VI_q$ and $ VIII$. All Lie subgroups corresponding to these subalgebras
act freely and transitively on the $AdS_3$ space, except for the $ \pmb{VI_0} $ Lie group.
We have investigated
the spacetime structure of T-dual findings for $ AdS_3 $ background and also some their physical interpretations
by introducing some convenient coordinate transformations.
Among the four dual backgrounds to $ AdS_3 $, only the dual metrics constructed out on the semi-Abelian Drinfeld doubles
$ (VI_q,3\mathcal{A}_1)$ and  $ (VIII_{.i},3\mathcal{A}_1)$ have true singularities.

Most importantly, in the absence of $B$-field, for all metrics of $AdS_2$,  $AdS_2 \times S^1$, $AdS_3$, and also the metric of the
analytic continuation of $AdS_{2}$  we have investigated the conformal invariance
conditions of the backgrounds up to three-loop order.
The results render the $\alpha'$ expansion is uncontrollable,
and thus one can't guarantee the conformal invariance of the backgrounds.
Notice that in the case of these spacetimes we have applied the usual rules of non-Abelian T-duality
without further corrections.
We have obtained  the non-Abelian T-duals of the metrics of $AdS$ families by using the PL T-duality approach
and have then checked the conformal invariance conditions of the duals up to two-loop order.
Unfortunately,  all of the dual backgrounds corresponding to these metrics
do not remain conformally invariant up to two-loop order.
Indeed, this was expected, because we did not use the $\alpha'$-corrected rules
of non-abelian T-duality that are necessary to have conformal invariance
at two-loop order \cite{Wulff22}.
It has been shown that the PL duality can be extended to order
$\alpha'$, i.e. two loops in the $\sigma$-model perturbation theory,
provided that the map is corrected \cite{Wulff22} (see, also, \cite{{Hassler2},{Marques}}).
It is possible that one applies the usual rules of non-Abelian
T-duality without further corrections, and still be able to obtain two-loop
solutions (e.g. \cite{Godel}). However, in general, further corrections to the rules are
necessary. With the modified rules one will be able to find the right $\alpha'$-corrections to the non-Abelian dual backgrounds,
so that the two-loop equations are satisfied. We intend to address this problem in the future.

Finally, we have studied the Abelian T-duality of BTZ background up to $\alpha'$-corrections
by using the T-duality rules of KM, when the dualizing is implemented by the shift of directions $\varphi$ and $t$.
In dualizing on the direction $\varphi$, we have shown that the structure and asymptotic nature of the dual
spacetime including the horizons and singularity are the same as in charged
black string derived in \cite{Horowitz1} without $\alpha'$-corrections, whereas
in performing the duality with respect to the coordinate $t$ it has been found a
new three-dimensional black string for which we have determined the horizons and singularity.
For this case, we have also shown that the Abelian T-duality transformation of KM changes
the asymptotic behavior of solutions from $AdS_3$ to flat.

\subsection*{Acknowledgements}

This work has been supported by the research vice
chancellor of Azarbaijan Shahid Madani University under research fund No. 97/231.
%%%%%%%%%%%%%%%%%%%%%%%%%%%%%%%%%%%%%%%%%%%%%%%%%%%%%%%%%%%%%%%%%%%%%%%%%%%%%

\end{document}